\documentclass[a4paper,12pt]{article}
\pdfoutput=1 %

\usepackage{jheppub} 
\usepackage{amsmath,amssymb,graphicx,float,slashed,xcolor,multicol}
\usepackage{tabularx}
\usepackage{url}
\usepackage{footmisc}
\usepackage{amsfonts}
\usepackage{cancel}
\usepackage{color}
\usepackage{multirow} 
\usepackage{pifont}
\usepackage{epstopdf}
\usepackage{comment}
\usepackage{booktabs} 
\usepackage{natbib}
\usepackage{array}
\usepackage{mathrsfs}
\usepackage[toc,page]{appendix}
\usepackage{mathtools}
\usepackage{romannum}
\usepackage{ulem}
\usepackage{bbold}
\usepackage{enumitem}
\usepackage{multirow}
\usepackage{cleveref}
\usepackage{caption,subcaption}
\usepackage{float}
\usepackage{multirow}
\usepackage{upgreek}
\usepackage[toc,page]{appendix}
\usepackage{romannum}
\allowdisplaybreaks
\usepackage{bbm}                
\usepackage{xspace}				
\usepackage{tikz}
\usetikzlibrary{arrows,shapes}
\usetikzlibrary{trees}
\usetikzlibrary{matrix} 
\usetikzlibrary{positioning}				
\usetikzlibrary{calc,through}				
\usetikzlibrary{decorations.pathreplacing}  
\usepackage{pgffor}							
\usetikzlibrary{decorations.pathmorphing}	
\usetikzlibrary{decorations.markings}

\tikzstyle{block} = [draw, rectangle, 
minimum height=3em, minimum width=6em]

\newcommand*{\rom}[1]{\expandafter\@slowromancap\romannumeral #1@}

\newcommand{\eg}{\textit{e.g.}}

\def\beq{\begin{equation}}
\def\eeq{\end{equation}}
\def\bea{\begin{eqnarray}}
\def\eea{\end{eqnarray}}

\title{Dark photons and axion-like particles at the Electron-Ion Collider in China}
\author[a]{Qiyuan Gao,}
\author[a]{Dan Lin,}
\author[b,c]{Hongkai Liu,}
\author[a]{Teng Ma}
\affiliation[a]{ICTP-AP, 
University of Chinese Academy of Sciences, 100190 Beijing, China} 
\affiliation[b]{High Energy Theory Group, Physics Department,
Brookhaven National Laboratory, Upton, NY 11973, USA}
\affiliation[c]{Physics Department, Technion – Israel Institute of Technology, Haifa 3200003, Israel}

\abstract{ The Electron-Ion Collider in China (EicC), a proposed high-luminosity facility with advanced charged particle and photon detection capabilities, provides unique opportunities to uncover new physics beyond the Standard Model.
We analyze its sensitivity to dark photons produced through electron bremsstrahlung in coherent scattering. Thanks to its beam energy settings, it has the potential to comprehensively probe the previously unexplored parameter space between the constraints from meson decays and beam dumps below $\mathcal{O}(1)$ GeV  with displaced-vertex search. 
Additionally, the EicC has the potential to probe axion-like particles (ALPs) in the mass range \( 0.1 \, \text{GeV} \lesssim m_a \lesssim 5 \, \text{GeV} \), with a coupling reach of \( \Lambda \lesssim  10^6 \, \text{GeV} \) , by combining the prompt-decay and displaced-vertex searches. The projected sensitivities to ALPs exceed the current bounds.
}
\begin{document}
\titlepage	
\maketitle

\flushbottom

\section{Introduction}
\label{sec:intor}

The discovery of the 125 GeV Higgs at LHC~\cite{ATLAS:2012yve,CMS:2012qbp} indicates the standard model (SM) precisely describes the particle physics below the TeV scale. However, there are a lot of puzzles that cannot be explained by SM, such as the Higgs hierarchy problem~\cite{tHooft:1979rat}, dark matter~\cite{Zwicky:1933gu}, and quantum chromodynamics (QCD) strong CP problem~\cite{Jackiw:1975fn,Peccei:1977hh}. In order to solve these problems, new physics beyond SM~(BSM) generally should be introduced. 

One type of new physics is the feebly interacting particles which interact extremely weakly with SM particles, making their detection a formidable experimental and theoretical challenge. Their potential to address unresolved problems in cosmology and particle physics has driven significant interest. Among these, the dark photons and ALPs are particularly intriguing and have been studied extensively.          

The dark photon associated with a dark $U(1)_{\text{D}}$ gauge, is predicted in many BSMs, such as the neutral naturalness mechanisms~\cite{Chacko:2005pe,Burdman:2006tz,Craig:2014aea,Csaki:2017jby,Geller:2014kta,Low:2015nqa,Barbieri:2015lqa,Csaki:2019qgb} and the dark matter (DM) models~\cite{Arias:2012az,Delaunay:2020vdb}. Unlike the photon, the dark photon is supposed to primarily couple to particles in the dark sector. But it can be designed to interact with SM particles through its kinetic mixing with the photon~\cite{Holdom:1985ag} or directly assigning  $U(1)_{\text{D}}$ charge to SM fields~\cite{Galison:1983pa}, providing a non-gravitational window into the dark sector.  Besides this, the dark photon as a mediator between dark sector and SM sector can be used to explain the anomalous excesses in cosmic rays~\cite{Pospelov:2008jd,Arkani-Hamed:2008hhe}.

The ALP is the hypothetical pseudo-scalar particle that  extends the concept of the axion, originally introduced to solve the QCD strong CP problem ~\cite{Peccei:1977hh,Peccei:1977ur}. Unlike the QCD axion, ALPs do not have a strict relationship between their mass and coupling strength, making them versatile candidates for a wide range of phenomena in cosmology, astrophysics, and particle physics. For example, its feeble coupling to SM field makes it a good dark matter candidate~\cite{Preskill:1982cy,Abbott:1982af,Dine:1982ah}, and ALPs can also significantly influence the evolution of the early universe~\cite{Peccei:2006as,Marsh:2015xka}.   

The Electron-Ion Collider (EIC) is a powerful machine to explore many kinds of new physics, as demonstrated by, \eg~\cite{Gonderinger:2010yn, Cirigliano:2021img,Davoudiasl:2021mjy,Zhang:2022zuz,Batell:2022ogj,Yan:2022npz,Davoudiasl:2023pkq,Boughezal:2022pmb,Liu:2021lan,Balkin:2023gya,Wang:2024zns,Davoudiasl:2024vje,Wen:2024cfu}. The EicC~\cite{Anderle:2021wcy, EicC} is a proposed next-generation high-intensity lepton-ion collider, focusing on different kinematics and perspectives compared to the EIC in the US~\cite{AbdulKhalek:2021gbh}. The energy of the electron beam is 3.5~GeV, and that of the ion beam is 20 GeV per nucleon. In this work, we explore the potential of the EicC to detect dark photons and ALPs via coherent scattering between electron and lead.

Dark photons can be produced via the electron bremsstrahlung copiously both at the EicC and EIC, benefiting from the $Z^2$ enhancement in coherent scattering, where $Z$ represents the atomic number of the ion. However, due to the high energy of the electron beam at the EIC, the dark photons are highly boosted and collinear with the initial electrons, making their detection challenging~\cite{Balkin:2023gya}. With a lower electron beam energy, we find that the EicC can probe previously unexplored regions of the dark photon parameter space, covering a mass range of $[2m_e, 3 ,\text{GeV}]$ and a coupling range to electromagnetic current of $[10^{-6}, 10^{-3}]$, thereby bridging the gap between the constraints from meson decays and beam dumps.  
 
For the ALP, it is produced through photon fusion at the EicC. We consider both prompt and displaced decays and find the EicC can probe the untouched parameter space below masses of $5$ GeV and the effective coupling of $1/\Lambda \sim 10^{-6}$ GeV$^{-1}$. For the displaced search, the EicC can better detect the smaller coupling region due to its smaller boost, compared to EIC.                    

The  paper is organized as follows. In Sec.~\ref{sec2}, we discuss the dark photon search at the EicC. In Sec.~\ref{sec3}, we study the ALP search and the conclusions are summarized in Sec.~\ref{sec4}. In the appendices, the details of the dark photon cross section calculations and the  numerical simulations are presented.      

\section{Dark photon}
\label{sec2}
%
\begin{figure}[ht]
    \centering
       \begin{minipage}{0.5\textwidth}
        \centering
        \includegraphics[width=\textwidth]{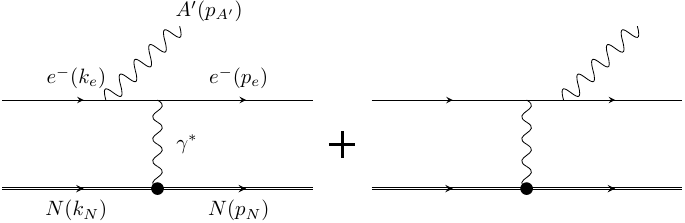}
    \end{minipage}\hfill
    \begin{minipage}{0.45\textwidth}
        \centering
        \includegraphics[width=\textwidth]{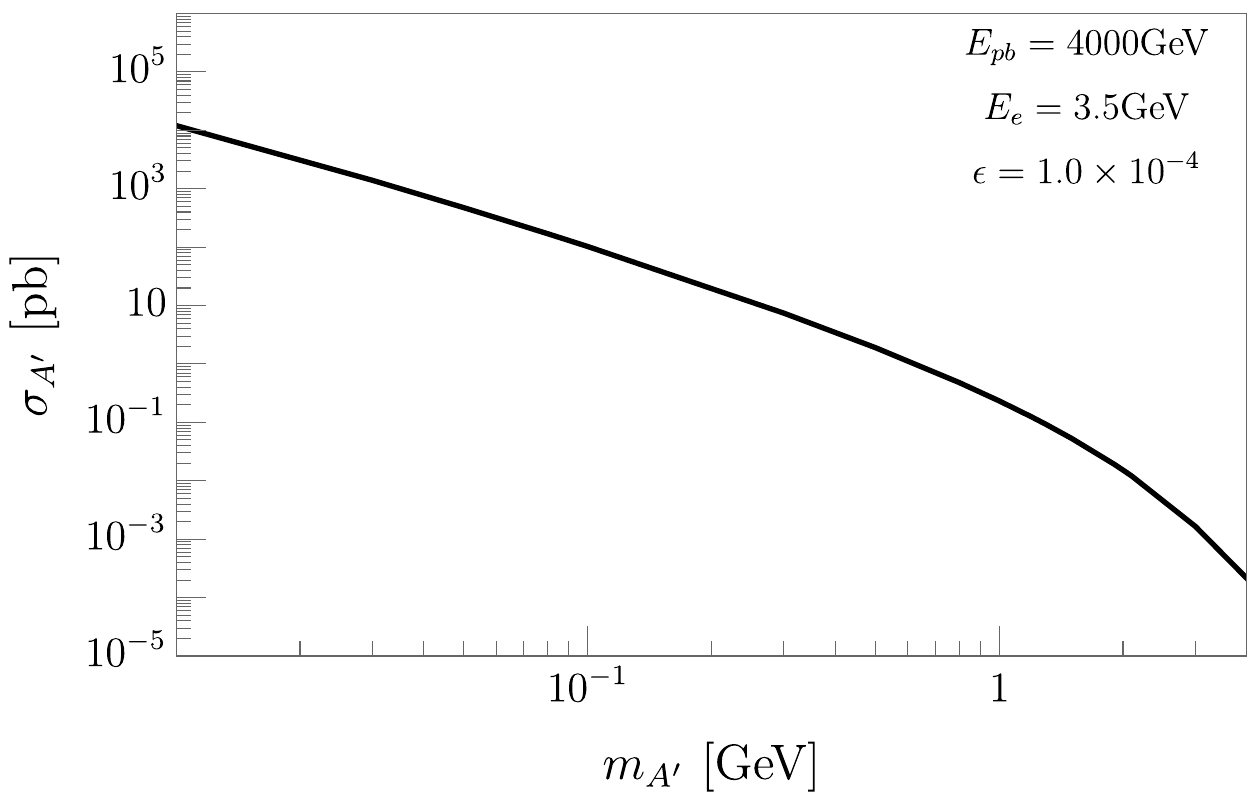}
    \end{minipage}
    \caption{ Left panel: Feynman diagram of dark photon production via electron bremsstrahlung at the EicC. Right panel: the cross section of dark photon production via the electron bremsstrahlung with  $\epsilon=10^{-4}$.  
    }
    \label{fig:Feyn_dp}
\end{figure}
We begin with the coherent production of dark photons. We consider the minimal dark photon model that the dark photon $A^\prime$ interacts with SM fields via the kinematic mixing with photon, 
\begin{align}
    \mathcal{L}_{A'} = -\frac14 F^{\mu\nu}F_{\mu\nu}-\frac14 F^{'\mu\nu}F'_{\mu\nu}-\frac{\epsilon}{2}F^{\mu\nu}F'_{\mu\nu}+\frac12 m_{A'}^2 {A'}^2+e A_\mu J^\mu_{\text{\tiny em}}\,,
\end{align}
where $ F^{\prime \mu\nu}$ is the dark photon field strength. Working on the mass eigenstates, the physical dark photon $A^\prime$ with massive $m_{A^\prime}$ couples to the SM particles with strength proportional to the mixing parameter $\epsilon$ and the interaction can be described by the Lagrangian
\beq
\mathcal{L}^{\rm int}_{A'} = -e\epsilon A^{\prime}_{\mu}J^\mu_{\text{em}},
\eeq
where $J^\mu_{\text{em}}$ is the SM electromagnetic current. The dark photons can be produced through the electron bremsstrahlung at the EicC, $ e\left(k_e\right)+N\left(k_N\right) \rightarrow e\left(p_e\right)+N\left(p_N\right)+A^{\prime}\left(p_{A^{\prime}}\right)$, as shown in the left panel of Fig.~\ref{fig:Feyn_dp}.  The dark photon can also be produced through ion bremsstrahlung, which can enhance its production. However, the generated dark photon will be highly boosted and nearly colinear, making it challenging to detect the final-state leptons. Therefore, to be conservative, we do not consider this contribution in our analysis. For the coherent production,  the maximal mass of the dark photon can be estimated by comparing the momentum of the virtual photon with the ion radius, which determines the maximal momentum of virtual photon. In the limit $m_e$, $m_{A^{\prime}} \ll m_N \sim \sqrt{s}$, the minimum exchanged photon momentum can be estimated as  
\beq
Q_N^2 \simeq \frac{m^4_{A^\prime}}{m_N^2}\left(1-\frac{s}{m_N^2}\right)^{-2}\,.
\eeq
Coherent scattering is highly suppressed when the minimum momentum transfer $Q_N$ at the ion exceeds the inverse of the ion's radius, $r_N \sim A^{1/3}(1\,\text{fm})$. By requiring $Q_N < 1/r_N$, we can estimate that the maximum dark photon mass can be produced coherently 
\beq
\left(m_{A^{\prime}}\right)_{\max } \sim 7.7~  
\mathrm{GeV}\left(\frac{m_N}{193 \mathrm{GeV}}\right)^{-\frac{1}{2}}\left(\frac{\sqrt{s-m_N^2}}{237\mathrm{GeV}}\right)\left(\frac{A}{207}\right)^{-\frac{1}{6}}.
\eeq
The dark photon coherent production cross section is shown in the right panel of Fig~\ref{fig:Feyn_dp} with mixing parameter fixed at $\epsilon =10^{-4}$.
For more details on the calculation of the $e^{-} N \rightarrow e^{-} N A^{\prime}$  cross section, see Appendix~\ref{apppendB}.

\begin{figure}[ht]
    \centering
       \begin{minipage}{0.45\textwidth}
        \centering
        \includegraphics[width=\textwidth]{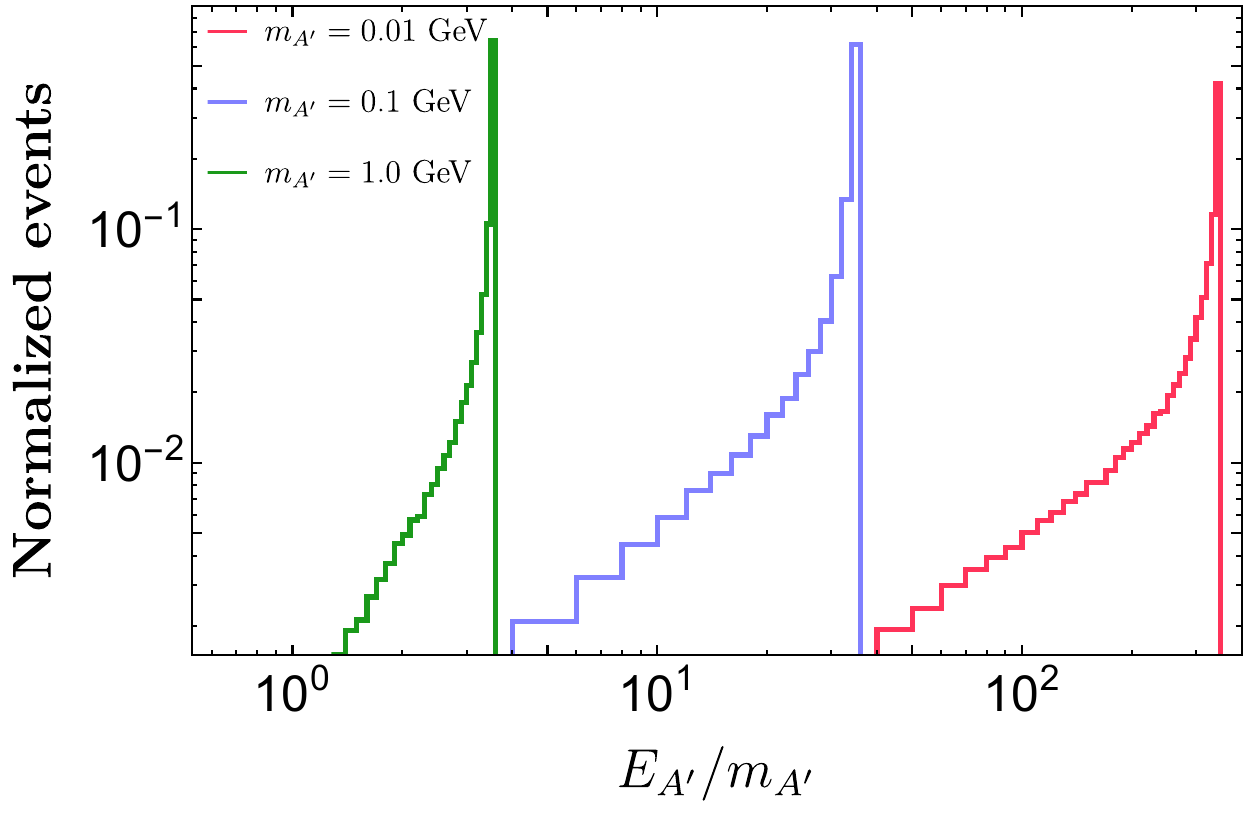}
    \end{minipage}\hfill
    \begin{minipage}{0.45\textwidth}
        \centering
        \includegraphics[width=\textwidth]{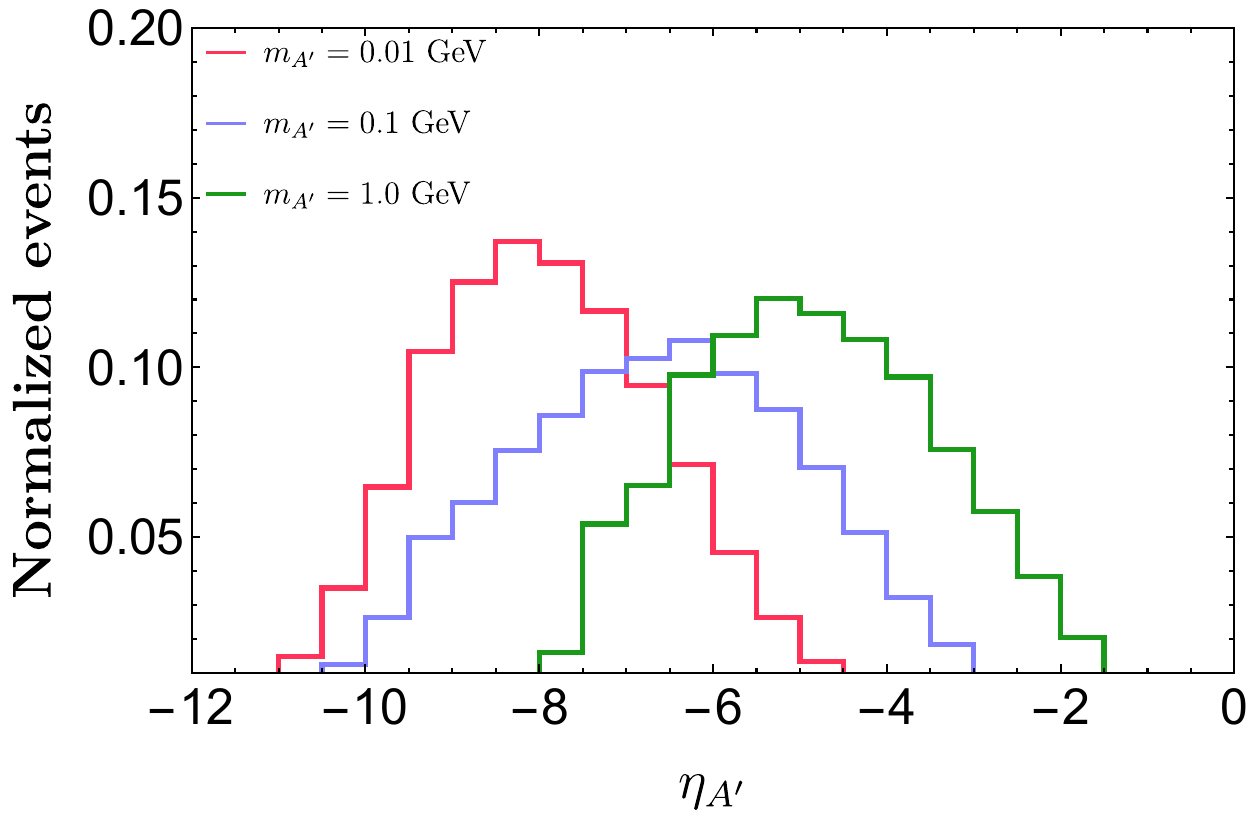}
    \end{minipage}
    \caption{  The normalized $E_{A^{\prime}}/m_{A^{\prime}}$~(left panel) and $\eta_{A^{\prime}}$~(right panel) distribution for dark photons with masses of 0.01 GeV (red), 0.1 GeV (blue), and 1 GeV (green).
 }
    \label{fig:8}
\end{figure}

In the bremsstrahlung process, the dark photon takes most of the electron's energy, resulting in energy distributions with a  peak around \( E_{A'} \sim 3.5 \, \text{GeV} \). Additionally, the dark photon aligns with the electron (backward direction) due to the bremsstrahlung process. These kinetic properties of the dark photon are consistent with our numerical simulations, as shown in Fig.~\ref{fig:8}. The left (right) panel of Fig.~\ref{fig:8} presents the boost factor $E_{A^{\prime}}/m_{A^{\prime}}$ (pseudorapidity   $\eta_{A^{\prime}}$) distributions for dark photons with masses of $\{0.01, 0.1, 1\}$ GeV, shown in red, blue, and green, respectively.

\subsection{Displaced-vertex search }
\begin{figure}[ht]
    \centering
    \begin{minipage}{0.45\textwidth}
        \centering
        \includegraphics[width=\textwidth]{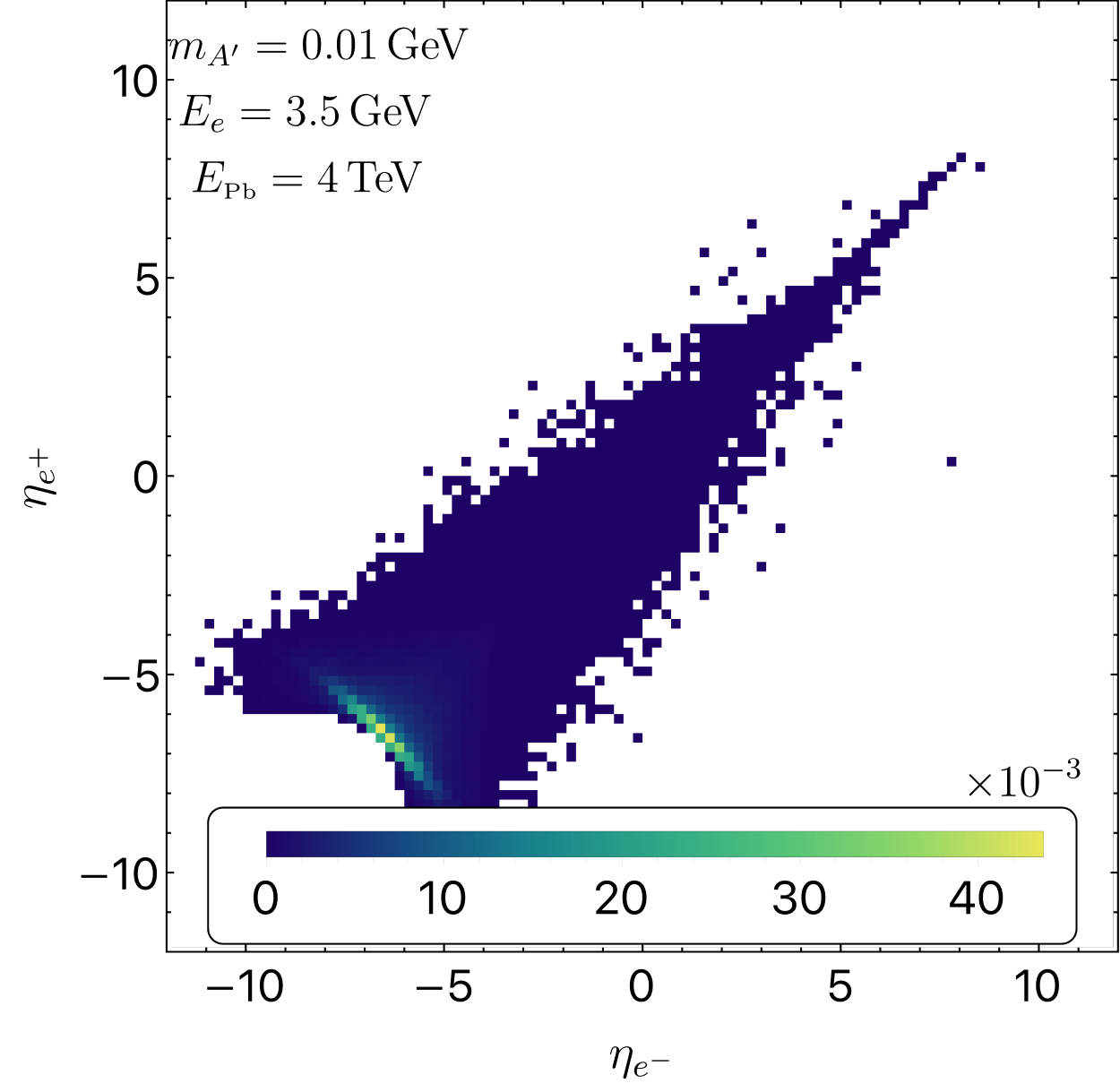}
        \end{minipage}\hfill
    \begin{minipage}{0.45\textwidth}
        \centering
        \includegraphics[width=\textwidth]{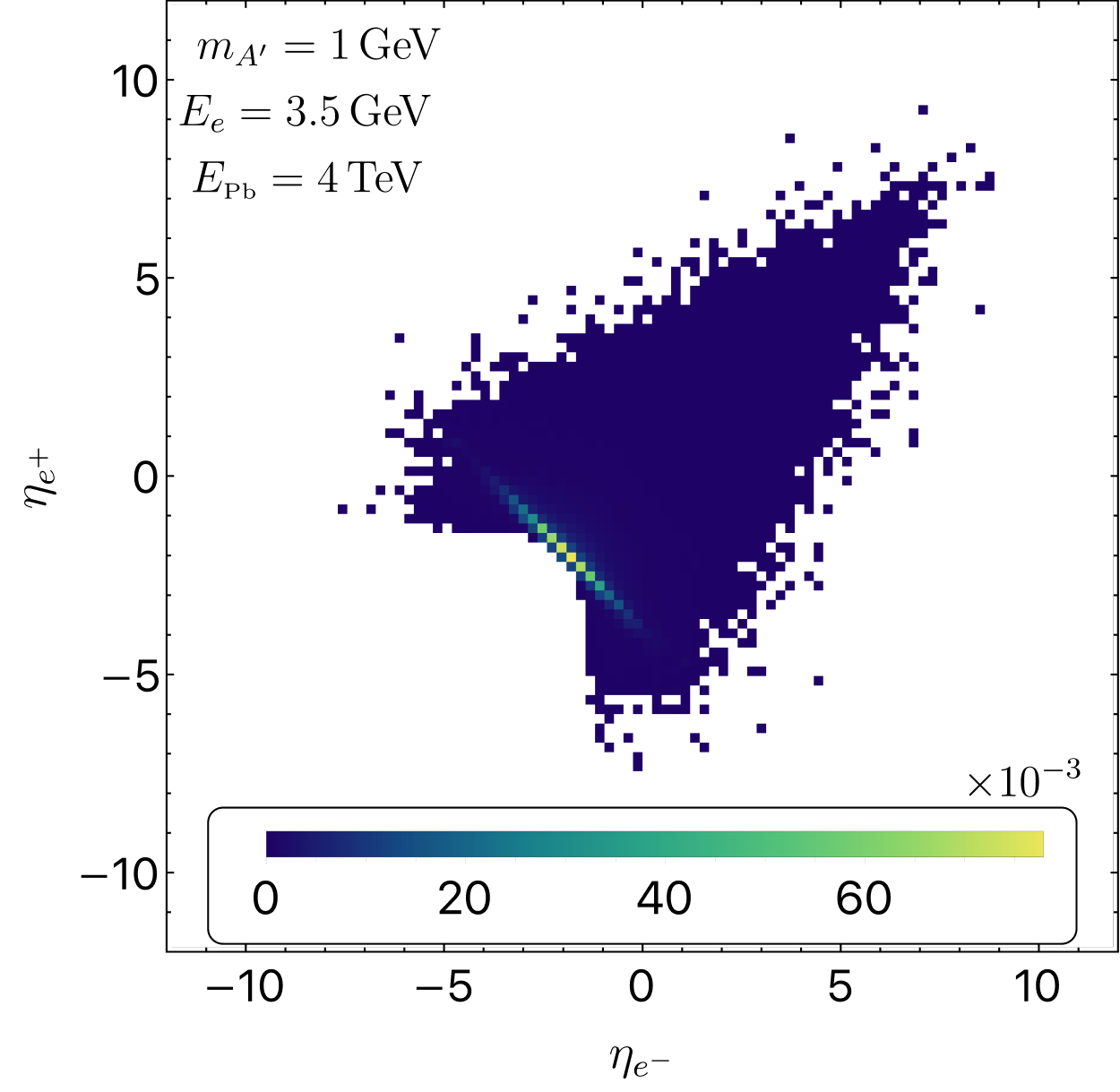}
    \end{minipage}
    \caption{The pseudorapidity distribution of the final-state lepton pair from dark photon decays with masses 0.01~GeV~(left panel) and 1 GeV~(right panel). }
    \label{fig:9}
\end{figure}

\begin{figure}
    \centering
    \includegraphics[width=5.5cm, height=5.5cm]{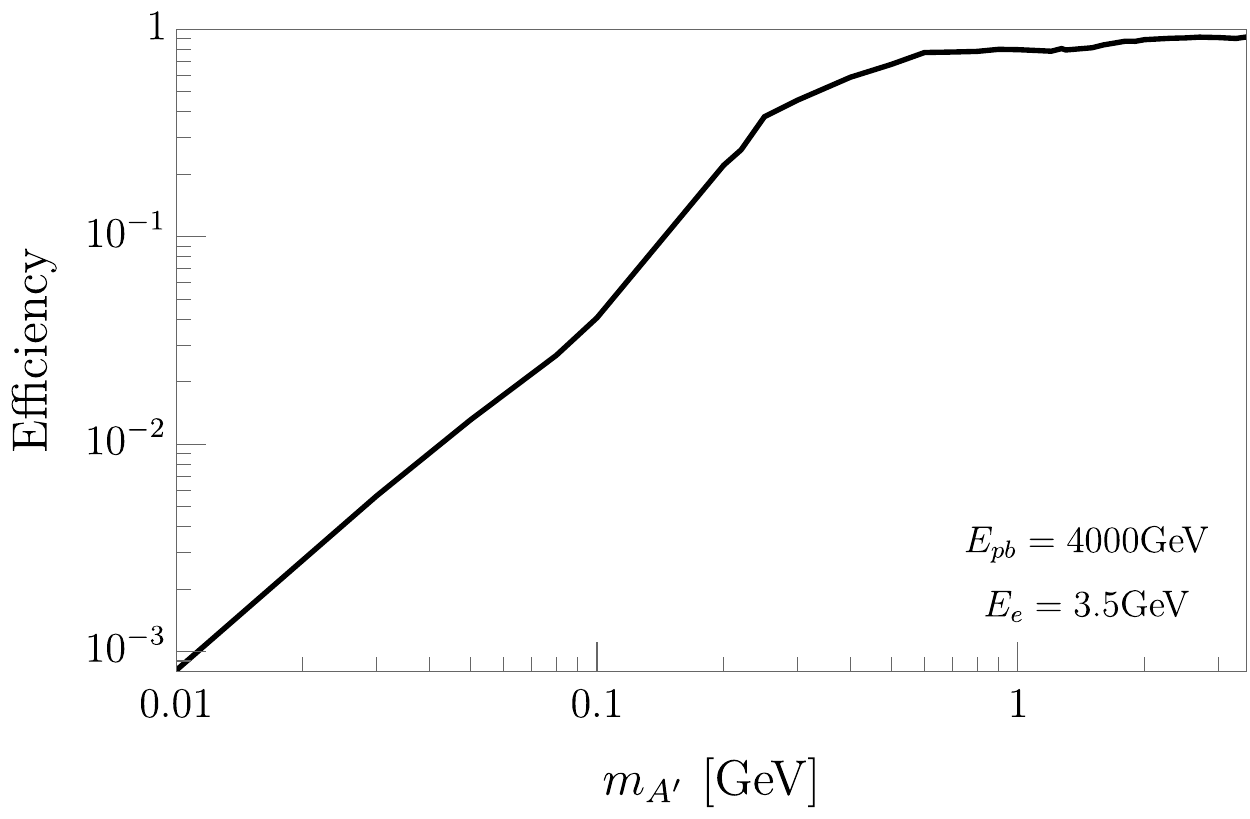}
    \includegraphics[width=0.4\linewidth]{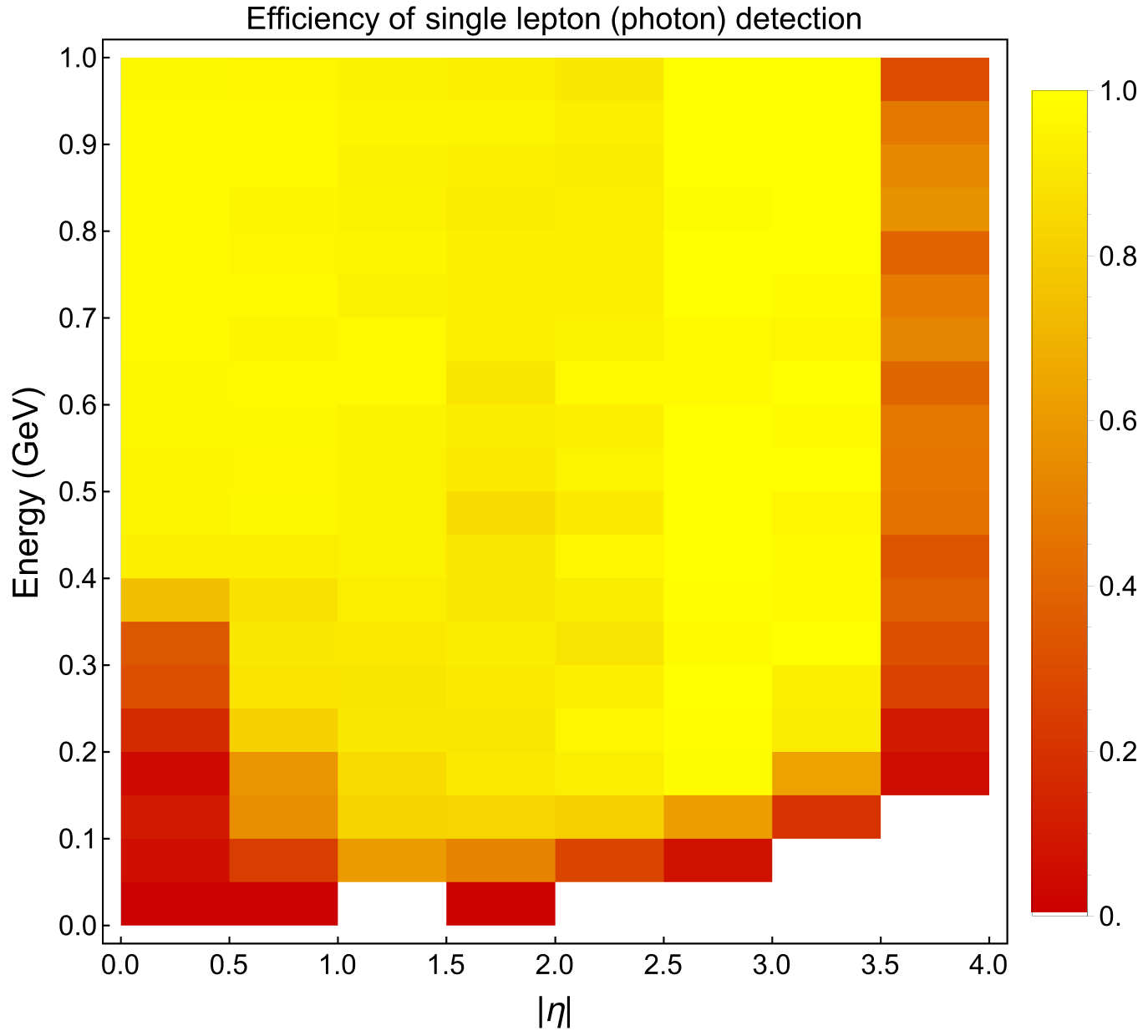}
    \caption{Left panel: the detecting efficiency of the electrons after applying the cuts in Eq.~\ref{eq.3.8}. Right panel: the efficiency of detecting a single lepton or photon with a specific energy and pseudorapidity.}
    \label{fig:eff}
\end{figure}

\begin{figure}[ht]
    \centering
       \begin{minipage}{0.48\textwidth}
        \centering
        \includegraphics[width=\textwidth]{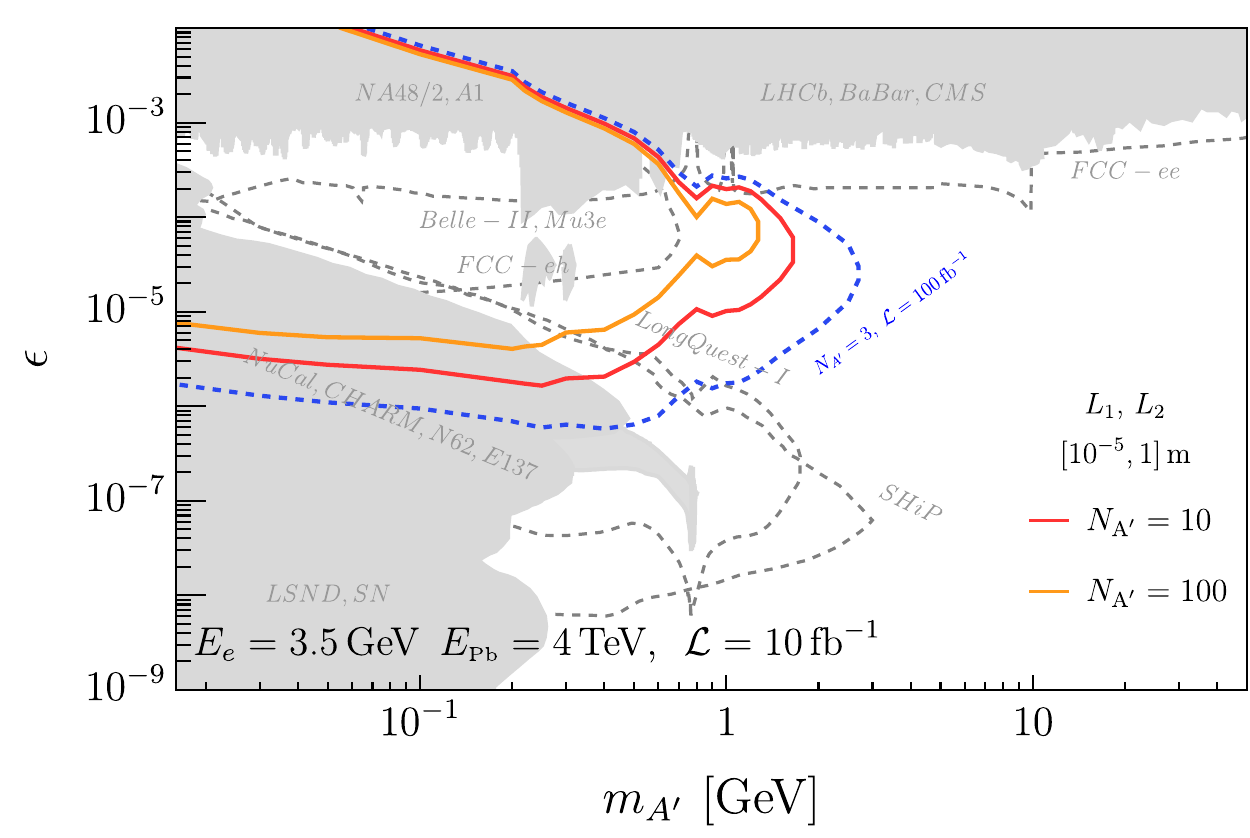}
    \end{minipage}\hfill
    \begin{minipage}{0.48\textwidth}
        \centering
        \includegraphics[width=\textwidth]{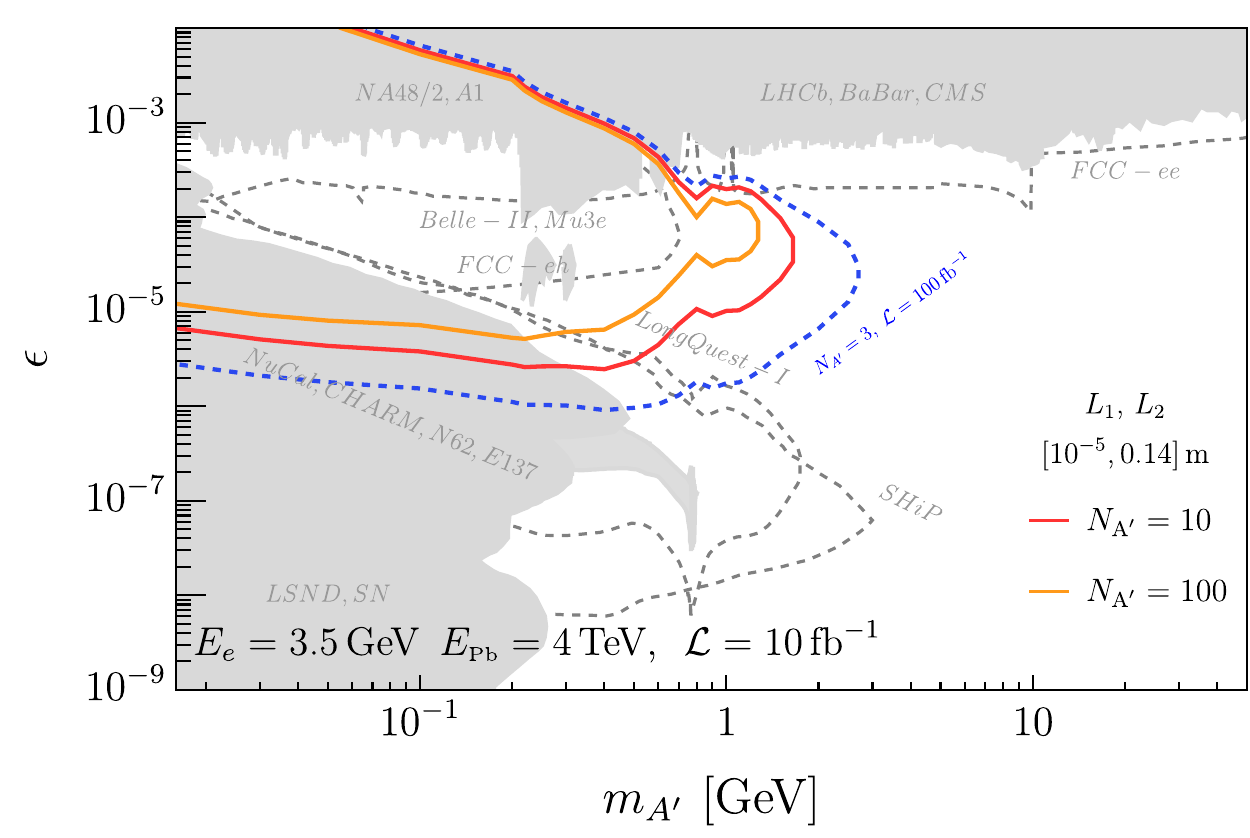}
    \end{minipage}
    \caption{
   The projected sensitivity of the EicC for displaced dark photon searches is shown, with $E_e = 3.5 \, \mathrm{GeV}$, $E_{pb} = 4 \, \mathrm{TeV}$, and an integrated luminosity of $\mathcal{L} = 10 \, \mathrm{fb}^{-1}$. Sensitivities corresponding to $N_{A^{\prime}} = 10$, $100$  are represented by red and orange respectively. Considering the improvement of the facility, we plot the sensitivity defined by \( N_{A^{\prime}} = 3 \) and \( \mathcal{L} = 100 \, \mathrm{fb}^{-1} \), which is represented by the blue dashed line. The parameters $L_1$ and $L_2$ denote the minimum distance for reconstructing the charged particle pair production vertex and the decay volume, respectively. \(\{ L_1, L_2 \} = \{ 10 \, \mu\text{m}, 1 \, \mathrm{m} \, (\text{or} \, 0.14 \, \mathrm{m}) \}\)
. The gray shaded regions denote existing experimental constraints, while the dashed gray lines represent the projected sensitivities of proposed searches. }
    \label{fig:10}
\end{figure}

Due to its coupling with the electromagnetic current, the dark photon can decay to all kinematically accessible light charged states in SM. For masses within the range \( 2 m_e \lesssim m_{A^{\prime}} \lesssim 0.8 \) GeV, the dominant decay mode of the dark photon is into lepton pairs, with a decay rate given by \cite{Ilten:2016tkc}.
\begin{equation}
    \Gamma_{A^{\prime} \rightarrow \ell^{+} \ell^{-}}=\frac{\epsilon^2 \alpha}{3} m_{A^{\prime}}\left(1+2 \frac{m_\ell^2}{m_{A^{\prime}}^2}\right) \sqrt{1-4 \frac{m_\ell^2}{m_{A^{\prime}}^2}} 
\end{equation}
 where $\alpha$ is the fine-structure constant. To better understand the kinematical properties of the lepton pairs, the pseudo-rapidity distribution of electrons and positrons from dark photon decays is shown in Fig.~\ref{fig:9}. For the majority of events, the electrons travel along the direction of the initial electron beam because the dark photon is highly backward. To ensure that the charged leptons can be effectively detected, we require the leptons to satisfy the condition~\cite{EicC}
\begin{equation}
     \left|\eta_{\ell}\right| < 4.0,\quad E_{\ell} > 0.05~\mathrm{GeV}
     \label{eq.3.8}.
\end{equation}
The efficiency of those cuts is shown in the left panel of Fig.~\ref{fig:eff}. We can see that the efficiency is larger than 10\% for $m_{A^\prime} \gtrsim 0.1$~GeV.    
The efficiency of single-lepton detection depends on both energy and pseudorapidity, with an average value of $f(E, \eta) \approx 0.9$. The numerical simulation of the efficiency distribution is shown in the right panel of Fig.~\ref{fig:eff}. 
  Heavier dark photons can decay into various hadronic states, predominantly into $\pi^+ \pi^-$, with decay widths and branching ratios detailed in \cite{FASER:2018eoc,Araki:2020wkq}.
  Since particles from dark photon decays are charged, the decay vertices can be effectively reconstructed at the EicC, enhancing the feasibility of dark photon detection through precise vertex reconstruction. For a conservative estimate, we consider only lepton pairs as the signal in the displaced vertex search. The possibility for a dark photon decaying between distance $L_1$ and $L_2$ is
\begin{equation}
    P = e^{-L _1/ L_{A^{\prime}}} - e^{-L_2 / L_{A^{\prime}}}
\end{equation}
where $L_{A^{\prime}}=p_{A^\prime}/(m_{A^\prime}\Gamma_{A^{\prime}})$, $p_{A^\prime}$ is the magnitude of dark photon momentum in lab frame, and $\Gamma_{A^{\prime}}$  is the decay width. $L_1$ is the distance resolution of the charged particle vertex, which is approximately 10~$\mu$m and $L_2$ denotes the size of the detector~\cite{Anderle:2021wcy,EicC}. We consider two scenarios for the detector length scale. The first only involves the inner silicon barrel, with $L_2 = 14$~cm. The second includes the whole tracking system, taking $L_2 = 1$~m. 
The sensitivity of the EicC to dark photon parameter space depends on assumptions about the background. 
For illustration, we present dark photon sensitivity with two choices of signal event: $N_{A^\prime} = 10$, $100$, assuming a reasonable integrated luminosity of $10 \, \mathrm{fb}^{-1}$. The sensitivity for $L_2 = 1 \, \mathrm{m}$ and $L_2 = 0.14 \, \mathrm{m}$ is shown in the left and right panels of Fig.~\ref{fig:10}, respectively, with $N_{A^\prime} = 10$, $100$ represented by red and orange curves. In addition, considering future upgrades to the EicC, we plot the sensitivity defined by \( N_{A^{\prime}} = 3 \) and an integrated luminosity of \( 100 \, \mathrm{fb}^{-1} \) to illustrate the potential of such a search strategy, as indicated by the blue dashed lines in the left and right panels of Fig.~\ref{fig:10}. Comparing with the sensitivity defined by different $L_2$ , we find that a larger $L_2$ leads to increased sensitivity for detecting dark photons with longer lifetimes, as expected. The EicC demonstrates significant improvement over existing and ongoing experiments, such as NA48/2\cite{NA482:2015wmo}, A1\cite{Merkel:2014avp}, LHCb\cite{LHCb:2019vmc}, BaBar\cite{BaBar:2014zli}, CMS\cite{CMS:2019kiy}, E137\cite{Batell:2014mga}, NuCal\cite{Blumlein:2013cua, Blumlein:2011mv, Tsai:2019buq}, CHARM\cite{Gninenko:2012eq, Tsai:2019buq}, NA62\cite{NA62:2023nhs, Tsai:2019buq}, LSND\cite{Magill:2018tbb} and SN\cite{Hardy:2016kme, Chang:2016ntp}. 
Compared to traditional beam dump experiments, the EicC has a higher center-of-mass energy (approximately 150 GeV electron beam in the ion's rest frame). 
Furthermore, the EicC requires much shorter decay length of the dark photon.  Therefore, the EicC can probe stronger dark photon coupling, surpassing the projections of future experiments, such as LongQuest-I\cite{Tsai:2019buq}, SHiP\cite{SHiP:2020vbd}, FCC-ee\cite{Karliner:2015tga}, Belle-II\cite{Ferber:2022ewf}, Mu3e\cite{Echenard:2014lma} and FCC-eh\cite{DOnofrio:2019dcp}.
Even if there exist some backgrounds, the bounds for signal event number $N_{A^\prime} =100$ can still uniquely cover large areas of unexplored parameter space, indicating the EicC is a good complement to other dark photon search experiments. 

\section{Axion-like particles}
\label{sec3}
\begin{figure}[ht]
    \centering
    \begin{minipage}{0.45\textwidth}
        \centering
        \includegraphics[width=\textwidth]{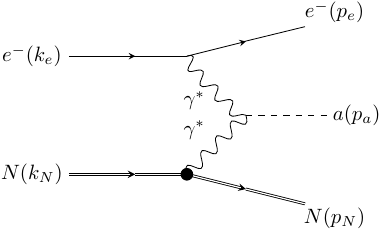}
    \end{minipage}\hfill
    \begin{minipage}{0.5\textwidth}
        \centering
        \includegraphics[width=\textwidth]{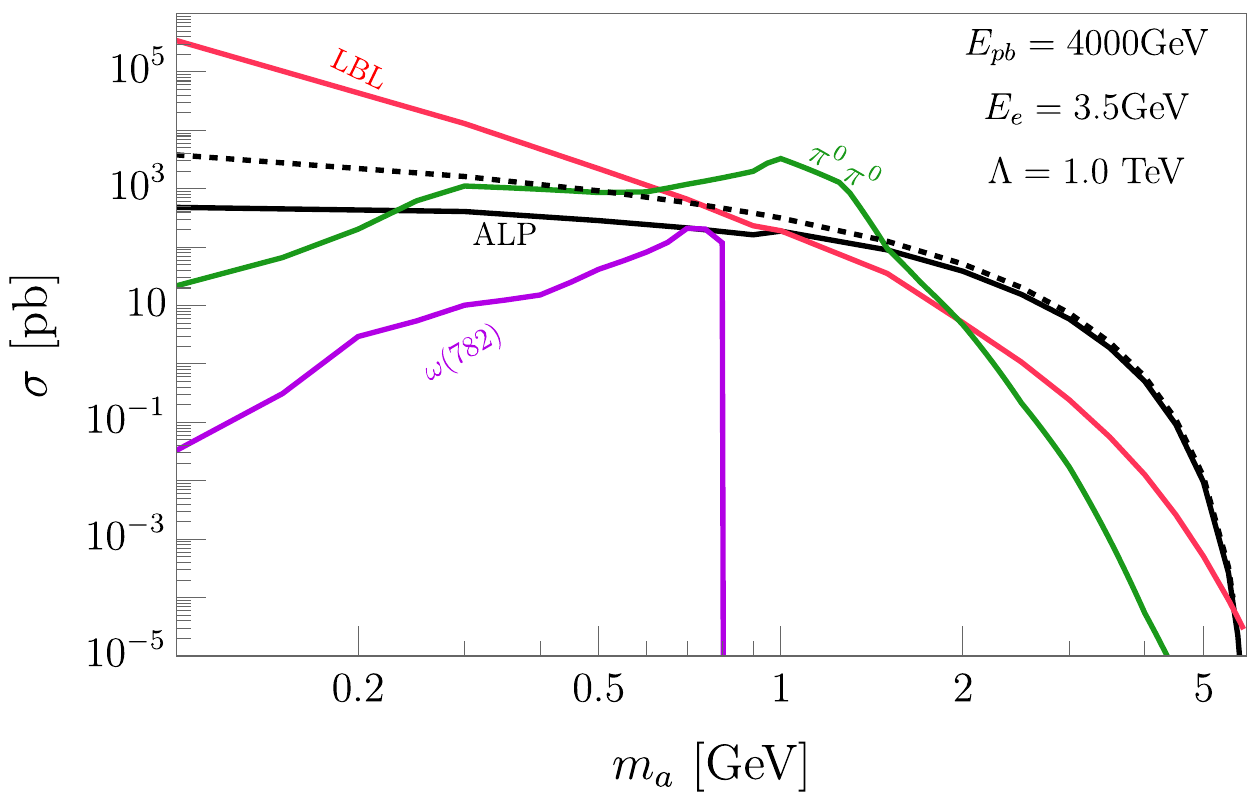}
    \end{minipage}
    \caption{Left panel: Feynman diagram of ALP production via the photon fusion at the EicC. Right panel: signal and background cross sections as a function of ALP mass $m_{a}$. The dashed and solid black lines represent the cross section for ALP production with \( \Lambda = 1 \, \text{TeV} \), calculated using the fully analytical method, before and after applying all cuts discussed in Sec.~\ref{sec:3.3}, respectively. The effective cross sections for the light-by-light(LBL), pion-pair, and omega backgrounds are shown by the red, green, and purple lines, respectively. The  background cross sections are $\sigma(m_a) \approx 4 \Delta m_{\gamma \gamma} \left( \frac{d \sigma}{d m_{\gamma \gamma}} \right) \Big|_{m_{\gamma\gamma}=m_a}
$ , where $\Delta m_{\gamma\gamma}$ is the invariant mass resolution as shown in Table~\ref{table:1}.}
    \label{fig:alp_diagram}
\end{figure}

In this section, we discuss the ALP coherent production at the EicC. We focus on the ALPs that mainly couple to photon,  
\begin{equation}
	\mathcal{L}_a \supset -\frac{1}{2}m_a^2a^2-\frac{a}{4\Lambda}F^{\mu\nu}\tilde{F}_{\mu\nu},
\end{equation}
where $\tilde{F}_{\mu\nu}\equiv (1/2)\epsilon_{\mu\nu\rho\sigma}F^{\rho\sigma}$ and $m_a$ denotes the mass of the ALP. Therefore, the ALPs can be produced via photon fusion at the EicC $e^- N\to e^-Na$ as shown in the left panel of Fig.~\ref{fig:alp_diagram}. Similar to the case of the dark photon, when the momentum transferred to the nucleus is smaller than the inverse of nucleus size, the electron interacts with the entire nucleus, resulting in a $Z^2$ enhanced cross section. In addition to its significantly larger cross-section, coherent scattering offers a relatively clean environment for the searches of ALPs. 
The EicC can probe the ALPs with mass up to $5~$GeV via the coherent scattering. 
The proton-proton and heavy-ion colliders at the LHC have already placed stringent constraints for $m_a\gtrsim5$~GeV. For ALPs below 0.1 GeV, it suffers from the strong bounds from the beam dumps experiments. Therefore, the specified target range for the EicC , spanning between 0.1 and 5 GeV, can fill the gap between the beam dumps and high-energy colliders.

The amplitude square of the process $e^{-} N \rightarrow e^{-} N a$ is 
\beq
\left|\mathcal{M}_a^{2 \rightarrow 3}\right|^2 \propto\left(Z^2 e^4\right) /\left(\Lambda_a^2 t_e^2 t_N^2 \right),
\eeq
where $t_e$ and $t_N$ represent the momentum transfer of the electron and ion, respectively. The total cross section as a function of ALP mass is shown by the dashed black curve in the right panel of Fig.~\ref{fig:alp_diagram}.
\begin{figure}[ht]
    \centering
    \begin{minipage}{0.45\textwidth}
        \centering
        \includegraphics[width=\textwidth]{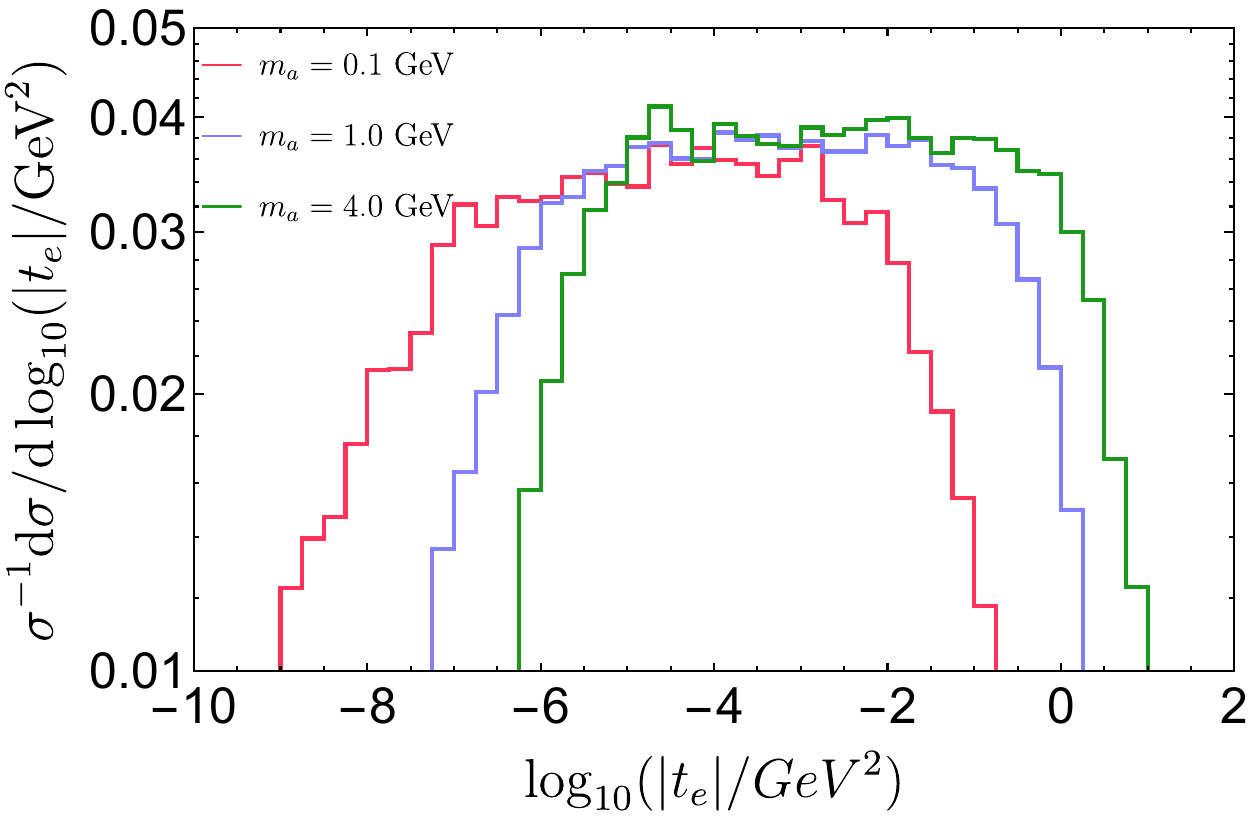}
    \end{minipage}\hfill
    \begin{minipage}{0.45\textwidth}
        \centering
        \includegraphics[width=\textwidth]{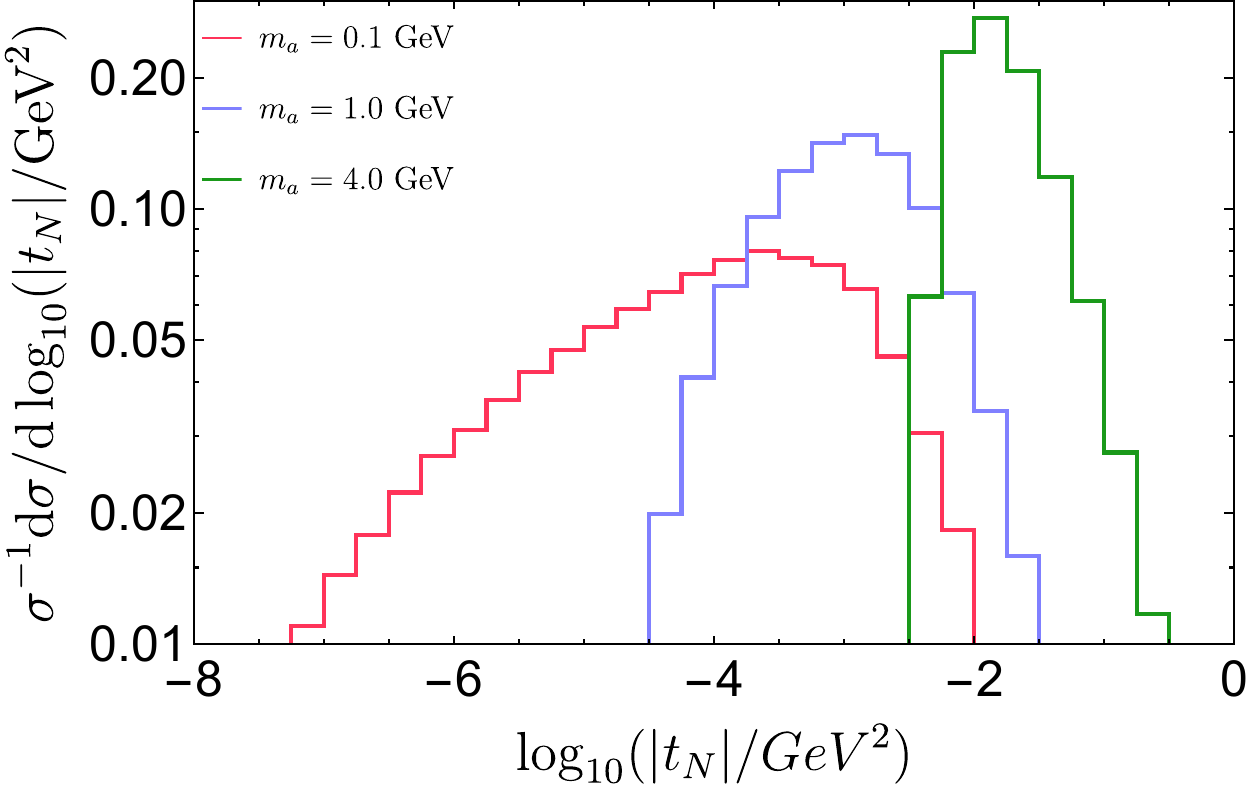}
    \end{minipage}
    \caption{Normalized $\mathrm{d} \sigma / \mathrm{d} \log \left|t_e\right|$ (left panel) and $\mathrm{d} \sigma / \mathrm{d} \log \left|t_N\right|$ (right panel) distributions of the ALP production cross section for ma = 0.1, 1.0 and 4.0 GeV plotted in red, blue and green, respectively.}
    \label{fig:2}
\end{figure}
Further details on the matrix element calculation are provided in Ref.~\cite{Balkin:2023gya}. 
In the limit $ m_a \ll \sqrt{s}$ , the minimal momentum transfer can be estimated as 
\begin{equation}  
\begin{aligned}
\left|t_e\right|_{\min } & \approx 1.24\times 10^{-11} \mathrm{GeV}^2\left(\frac{m_a}{1.0 \mathrm{GeV}}\right)^2\,,\, 
\left|t_N\right|_{\min } & \approx 1.19 \times 10^{-5} \mathrm{GeV}^2\left(\frac{m_a}{1 \mathrm{GeV}}\right)^4.
\end{aligned}
\end{equation}
In Fig. \ref{fig:2}, we plot the normalized differential cross section $\sigma^{-1}\mathrm{d} \sigma / \mathrm{d} \log |t|$ for $t=t_e $ (left-hand panel) and $t=t_N $ (right-hand panel) for $m_a$=\{0.1,1,4\} GeV. The cross section is dominated in the region where the momentum transfer is minimized, as expected. 
The normalized differential cross section  $\sigma^{-1}\mathrm{d} \sigma / \mathrm{d} \log |t_N|$ decreases rapidly at large momentum transfers due to the finite size of the nuclear. 

\subsection{prompt search}\label{sec:3.3}
\begin{figure}[ht]
    \centering
    \begin{minipage}{0.3\textwidth}
        \centering
        \includegraphics[width=\textwidth]{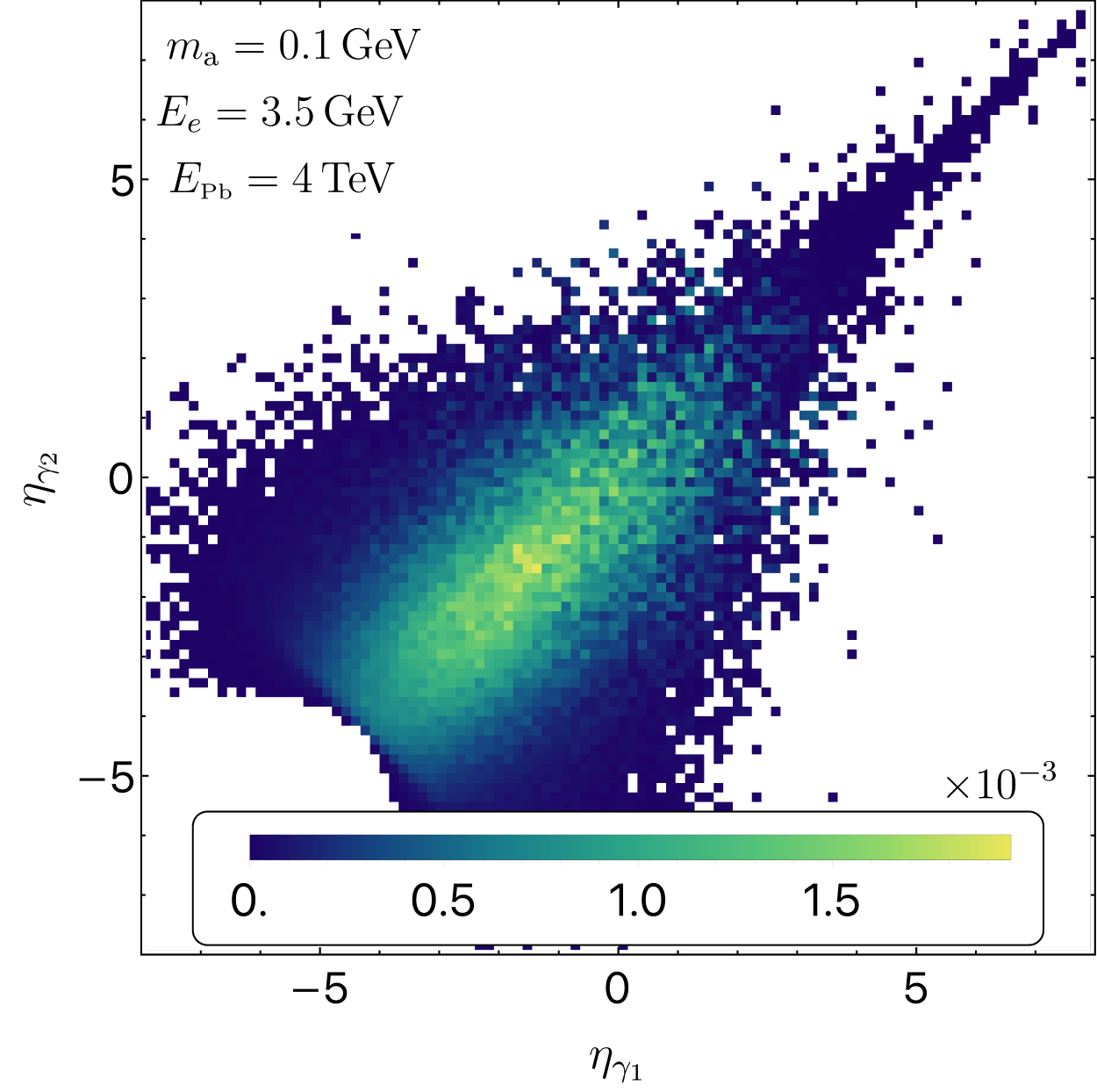}
        \end{minipage}\hfill
        \begin{minipage}{0.3\textwidth}
        \centering
        \includegraphics[width=\textwidth]{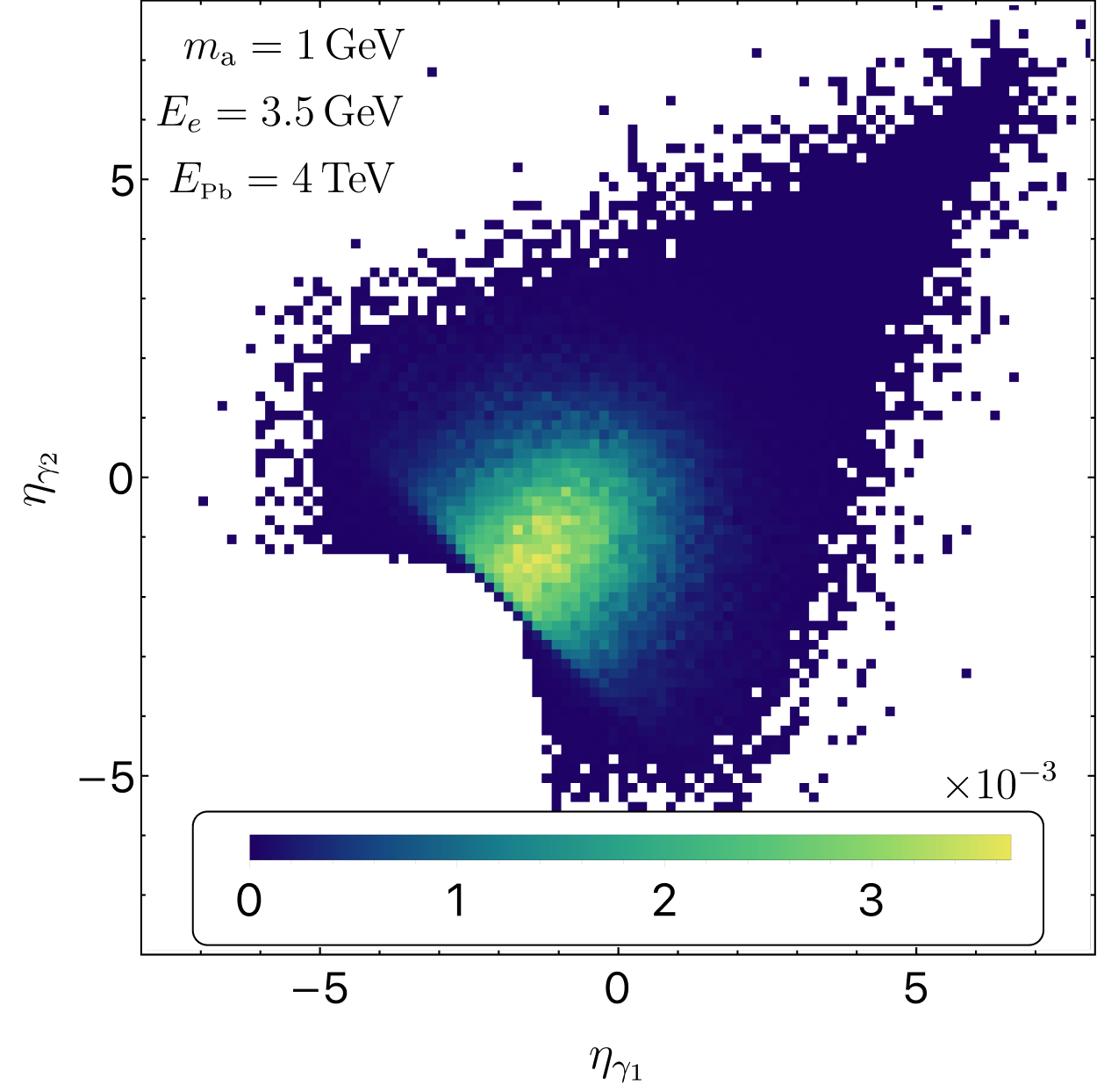}
        \end{minipage}\hfill
    \begin{minipage}{0.3\textwidth}
        \centering
        \includegraphics[width=\textwidth]{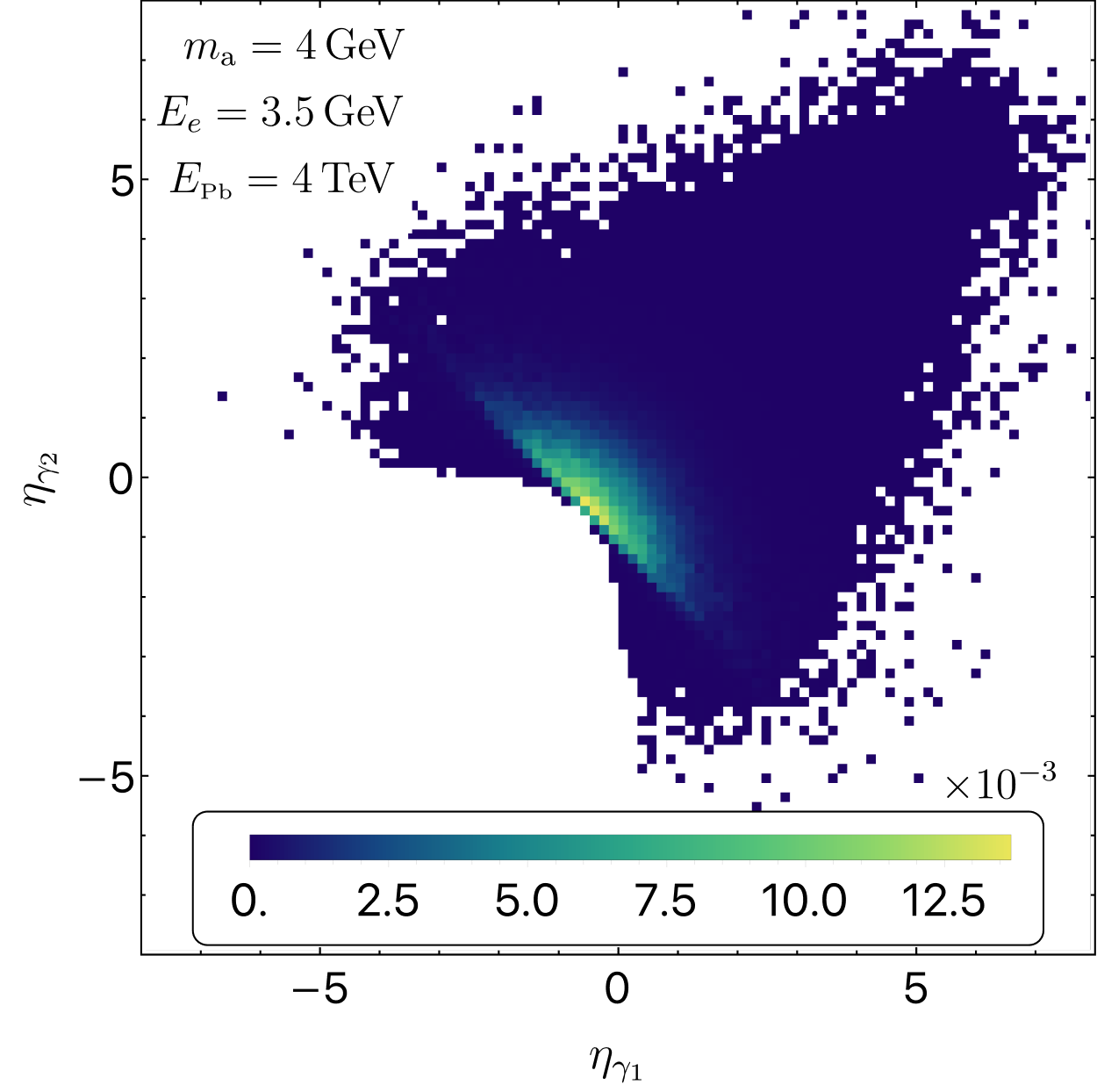}
    \end{minipage}
    \caption{The final diphoton pseudorapidity-dependent distribution from ALP decays is shown for three ALP masses: 0.1 GeV (left), 1 GeV (center), and 4 GeV (right). }
    \label{fig:4}
\end{figure}

The ALP decays back into photon pairs with the rate $\Gamma_{a \rightarrow \gamma \gamma}=m_a^3 /\left(64 \pi \Lambda^2\right)$. In the signal events, the final-state particles are scattered electron and ion, and photon-pair from the ALP decays. The scattered electrons and ions are very close to the beamline, which can easily escape from the central detector. Therefore, in this search, we only look for the di-photon in the final state. The pseudorapidity distribution of the signal photon is shown in Fig.~\ref{fig:4}. Due to the suppression by the ion form factor  at the large $|t_N|$ region, the photon distribution is concentrated in the negative pseudorapidity region. Photons from the decay of light ALPs have a broader pseudorapidity distribution.  This is because significant number of light ALPs can get most of energy from $t_N$, resulting in aligning with the ions.   By contrast, heavier ALPs are produced nearly at rest, resulting in a larger opening angle between two photons.   
Since soft and forward photons are challenging to detect, we require the following cuts for selecting the signals, 
\beq
E_\gamma > 0.05~\text{GeV},\quad\mid\eta_\gamma\mid < 4.0.
\label{eq:cut1}
\eeq
It is reasonable to assume that the efficiency of detecting a single photon is identical to the lepton detection efficiency shown in Fig.~\ref{fig:eff}. Our search strategy involves a bump hunt in the di-photon invariant mass spectrum near $m_a$
\beq
\mid m_{\gamma\gamma} - m_a\mid < 2 \Delta m_{\gamma\gamma},
\label{eq:cut2}
\eeq
where $\Delta m_{\gamma\gamma}$ represents the resolution of invariant mass of the two photons, which are summarized in Table~\ref{table:1}. The details are given in Appendix~\ref{appendixA}.

\begin{table}[h]
\centering
\begin{tabular}{cccccccccccc}
\toprule
$m_{\gamma\gamma}$ [GeV] & 0.1 & 0.3 & 0.5 & 0.7 & 0.9 & 1.0 & 2.0 & 4.0 & 7.0 & 10.0 & 15.0 \\ 
\midrule
$\Delta m_{\gamma\gamma} / m_{\gamma\gamma}$ & 0.123 & 0.062 & 0.046 & 0.042 & 0.035 & 0.033 & 0.023 & 0.018 & 0.016 & 0.016 & 0.013 \\
\bottomrule
\end{tabular}
\caption{The reconstructed ALP mass resolution $\Delta m_{\gamma \gamma} / m_{\gamma \gamma}$ for specific $m_{\gamma \gamma}$ values, determined by fitting the $m_{\gamma \gamma}$ spectrum and including the detector response and efficiency.}
\label{table:1}
\end{table}

The primary SM backgrounds are di-photon production via LBL scattering, $\pi_0$ pair production, and $\omega$ production. Their cross sections are estimated using the equivalent photon approximation (EPA)~\cite{Budnev:1975poe,Knapen:2016moh}, which assumes virtual photons behave as on-shell photons due to dominance in the region where $|t_{e/N}| \rightarrow 0$. Firstly, we discuss the main irreducible LBL background $e N \rightarrow e N \gamma \gamma$,
in which two virtual photons interact via one-loop box diagrams. Further details on the calculations of the LBL scattering cross section and simulations can be found in Refs.~\cite{Balkin:2023gya,Bern:2001dg}. 
Due to the finite coverage of the detector and inefficiencies in photon reconstruction, events where only two photons are detected among multiple produced photons are classified as background. We begin our examination with the production of neutral pion pairs, $e N \rightarrow e N \pi^0 \pi^0 ~\left(\pi^0 \pi^0 \rightarrow \gamma \gamma \gamma \gamma\right)$, where only two out of four photons are detected. The neutral pion pair production is influenced by light meson resonances in the low-$m_{\gamma\gamma}$ region~\cite{Klusek-Gawenda:2013rtu}, while the hand-bag model \cite{Diehl:2009yi} estimates the production rate in the high-$m_{\gamma\gamma}$ region. To estimate the cross-section of this background, we simulate $10^6$ events and select only those with exactly two photons in the final state based on the detection efficiencies shown in right panel of Fig.~\ref{fig:eff}. Additionally,  the production of the $\omega$ meson, which decays into three photons, $e N \rightarrow e N \omega \quad\left(\omega \rightarrow \pi^0 \gamma \rightarrow \gamma \gamma \gamma\right)$, also contributes to the background if one photon is missed. This background affects the region where $m_{\gamma\gamma} \lesssim m_\omega \approx 0.78 \mathrm{GeV}$. We use the results of $\omega$ photon production  $\gamma N \to \omega N$ in Ref. \cite{Ballam:1972eq} and EPA to estimate the cross section of $\omega$ production at the EicC. The branch ratio of $\omega \rightarrow \pi^0 \gamma$ is taken to be $8.28\%$ in the calculation.
\begin{figure}
    \centering
    \includegraphics[width=0.6\linewidth]{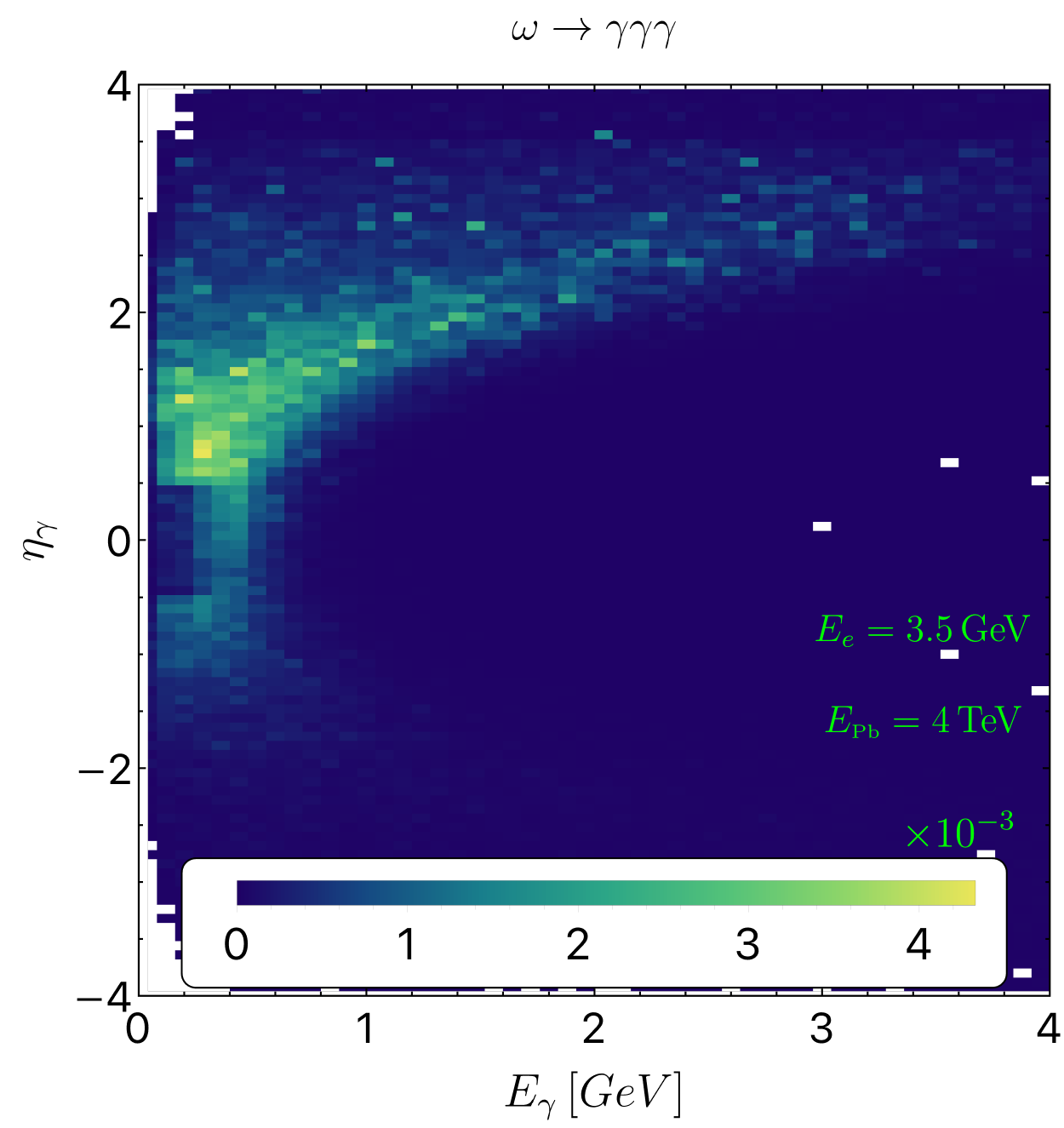}
    \caption{The energy and pseudorapidity distributions of the detected photons produced from the decay of $\omega$ meson.}
    \label{fig:omega decay}
\end{figure}
To reduce the reducible pion-pair production and $\omega$ production backgrounds, we further require that the detected two photons be back-to-back in the transverse plane:
\begin{equation}
\mid\Delta\varphi_\gamma -\pi \mid< 0.2
\label{eq:cut3}
\end{equation}
where $\Delta\varphi_{\gamma} =\varphi_{\gamma_1} - \varphi_{\gamma_2} $ is the difference in the azimuthal angle between the two detected photons.  Moreover, the detected photons produced from $\omega$ meson decay are more aligned with the ion, as indicated in  Fig. \ref{fig:omega decay}. Therefore, for $m_{\gamma\gamma} < m_\omega$, we require that the photons move along the direction of the incident electron
\begin{equation}
   m_{\gamma\gamma} < m_\omega:  -4 < \eta_{\gamma} < 0\,.
    \label{eq:cut4}
\end{equation}
Finally, the effective signal and background cross sections as a function of $m_a$, after applying the cuts in Eqs.~(\ref{eq:cut1} - \ref{eq:cut4}), are shown by the solid black and colored lines, respectively, in the right panel of Fig.~\ref{fig:alp_diagram}. 

\begin{figure}[ht]
    \centering
    \includegraphics[width=0.9\linewidth]{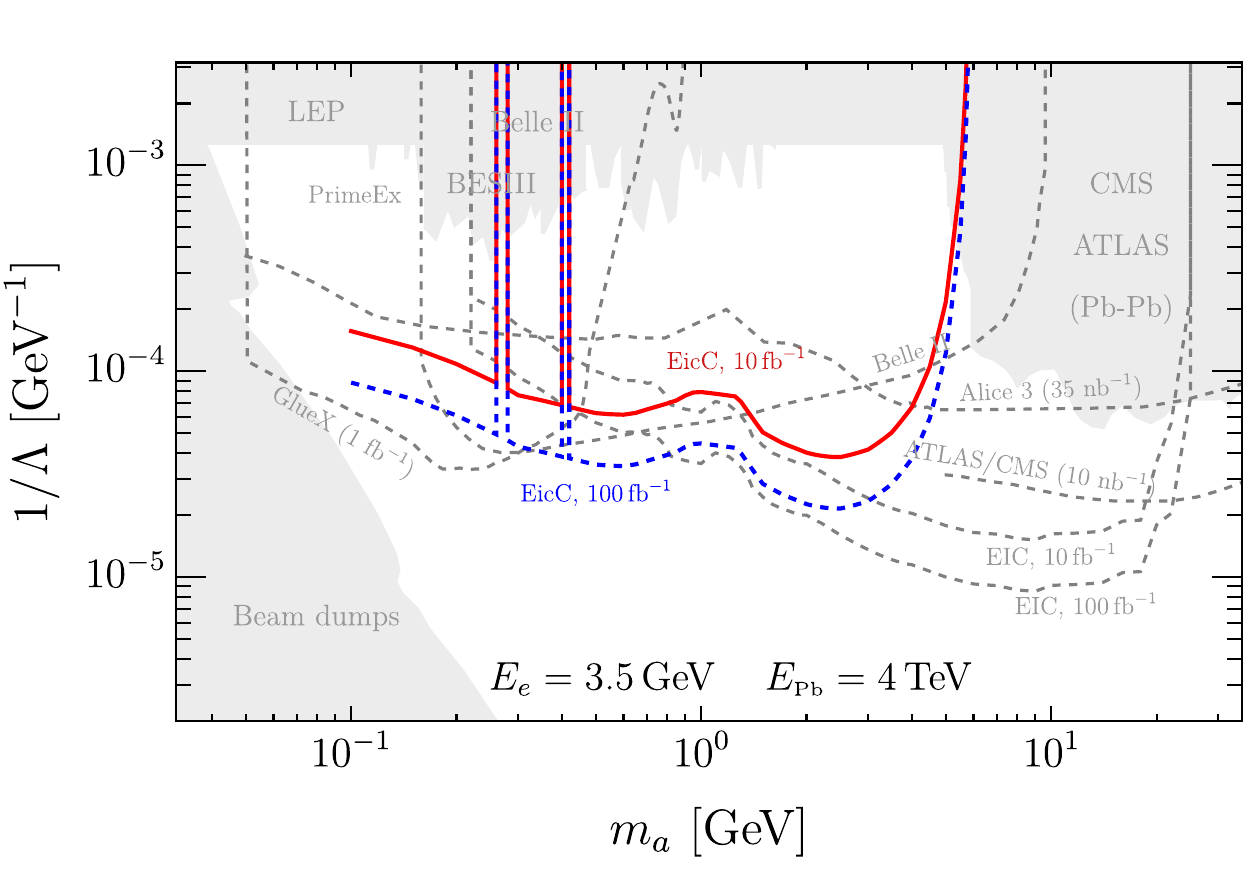}
    \caption{The projected sensitivity for ALP parameter space in prompt searches at the EicC. The red (blue dashed) line represents a benchmark luminosity of 10 (100) $\text{fb}^{-1}$. The gray shaded regions represent existing experimental constraints, and the dashed gray lines indicate the projected sensitivities of various proposed searches.
}
    \label{fig:5}
\end{figure}

The EicC sensitivity is estimated by requiring $S_a/\sqrt{B} = 2 $, where $S_a = \mathcal{L}\cdot\sigma_{\rm ALP}$ is the number of signal events. Here, $\mathcal{L} = 10~\text{fb}^{-1}$ and $100~\text{fb}^{-1}$ are the two benchmark values assumed for the integrated luminosity. Similarly, $B = \mathcal{L} \cdot(\sigma_{\rm LBL}+\sigma_{\pi^0\pi^0}+\sigma_{\omega})$ is the total number of background events. The EicC projected sensitivities in the ALP parameter space are shown in Fig.~\ref{fig:5}. Red (blue dashed) lines represent benchmark luminosities of 10 (100) $\text{fb}^{-1}$.
We can find that the EicC has the potential to explore previously uncharted parameter space with $1/\Lambda \ge 2 \times 10^{-5} \, \text{GeV}^{-1}$ and $0.1 \, \text{GeV} \lesssim m_a \lesssim 5 \, \text{GeV}$ for $\mathcal{L} =100\; \text{fb}^{-1}$.
The EicC demonstrates a significant improvement over existing bounds from experiments such as ATLAS/CMS, BESIII, Belle-II, LEP, and beam-dump experiments \cite{ATLAS:2020hii,Knapen:2016moh, Bauer:2017ris, OPAL:2002vhf, BESIII:2022rzz, Belle-II:2020jti, Jaeckel:2015jla, D_brich_2016, Bjorken:1988as} within the mass range $0.1 \, \text{GeV} \lesssim m_a \lesssim 5 \, \text{GeV}$.  Additionally, since the EicC detection can cover the future experiments, it can complement and provide cross-checks for these experiments such as GlueX \cite{Aloni:2019ruo}, Belle-II \cite{Ballam:1972eq}, heavy-ion programs \cite{Knapen:2016moh, dEnterria:2022sut}, and the EIC \cite{Balkin:2023gya}.

\subsection{Displaced-vertex search}
The EicC can provide information about the diphoton production vertex, enabling the reconstruction of the ALP decay vertex. Due to their weak couplings, ALPs typically have longer lifetimes than Standard Model particles that decay into di-photons. As a result, prompt Standard Model backgrounds can be effectively ignored in displaced vertex searches for long-lived ALPs. In the lab frame, the detected ALPs travel a distance $L_a$ larger than the diphoton vertex resolution, $L_R = 0.1 \, \text{m}$, but smaller than the distance to the EM calorimeter, $L_{\mathrm{EM}} = 1 \, \text{m}$. Similar to the dark photon case, the probability of the ALP decay within this volume is approximated by
\begin{equation}
P(L_R < L_a < L_{\mathrm{EM}}) \approx \exp \left(-\frac{L_R}{L_a}\right)-\exp \left(-\frac{L_{\mathrm{EM}}}{L_a}\right),
\end{equation}
where $L_a=p_a /\left(m_a \Gamma_{a \rightarrow \gamma \gamma}\right)$ and  $p_a$ is the ALPs' momentum in the lab frame. 

We consider a background-free search, and the estimated projections are evaluated by requiring \( N_a = 3 \). We also  require the photons that satisfy Eq.~(\ref{eq:cut1}) are detectable, and the single-photon detection efficiency is the same  as the prompt search, as shown in Fig.~\ref{fig:eff}.  The red and dashed blue lines represent the sensitivities with 10 and 100 $\text{fb}^{-1}$, respectively. The EicC demonstrates enhanced sensitivities for higher-mass ALPs compared to upcoming experiments like NA62 \cite{Dobrich:2019dxc}, FASER 2 \cite{Feng:2018pew}, SeaQuest \cite{Berlin:2018pwi}, and LUXE \cite{Bai:2021gbm}.

\begin{figure}[ht]
    \centering
    \includegraphics[width=0.9\linewidth]{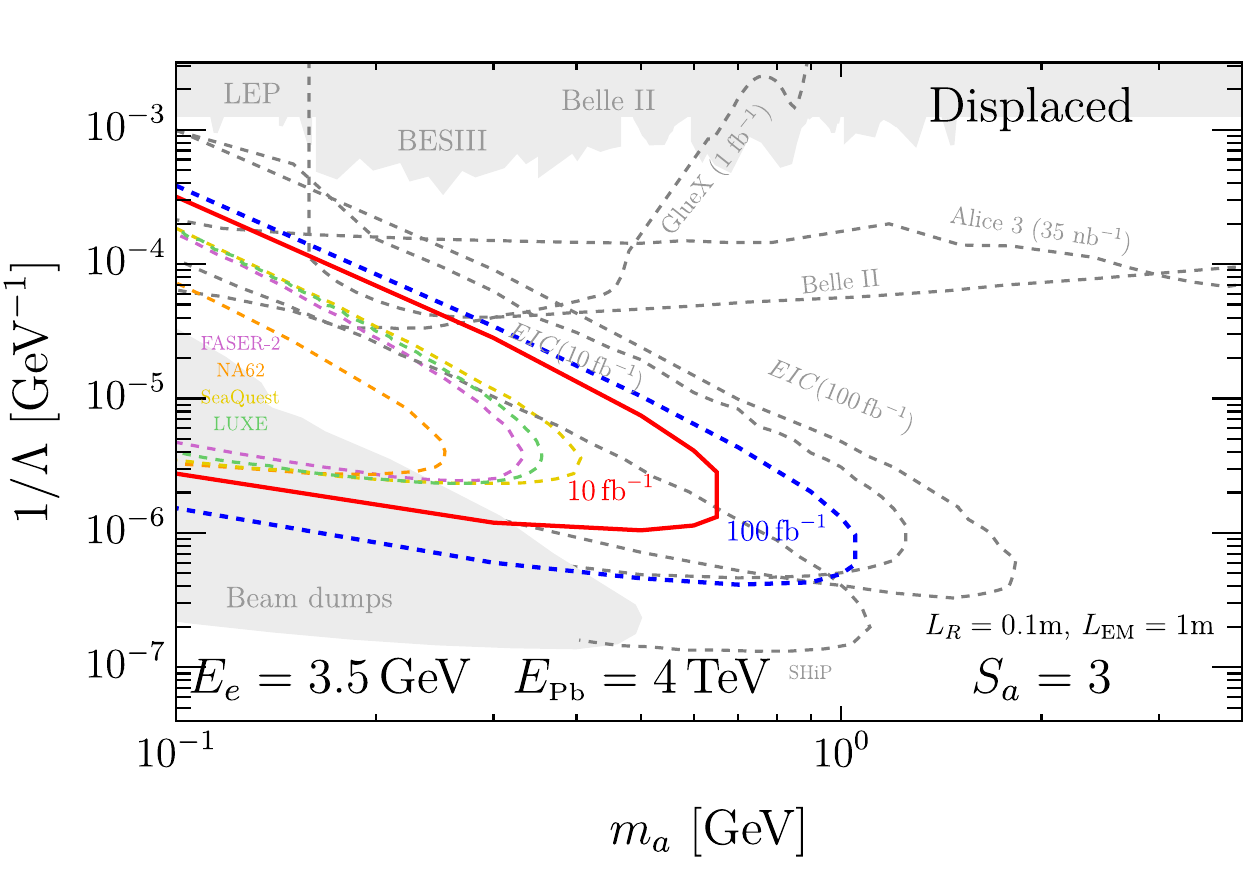}
    \caption{The projected sensitivity of the EicC from a displaced ALP search, with $E_e = 3.5$ GeV, $E_{pb} = 4$ TeV, and $\mathcal{L} = 10(100) \, \text{fb}^{-1}$, is defined by requiring $N_a = 3$. The red and dashed blue line represent the banchmark luminosity $\mathcal{L}=10\,\text{fb}^{-1}$ and $\mathcal{L}=100\,\text{fb}^{-1}$, respectively.  The diphoton vertex resolution is set to $L_R = 0.1 \, \text{m}$, and the distance from the interaction point to the EM calorimeter is $L_{\mathrm{EM}} = 1 \, \text{m}$.
}
    \label{fig:6}
\end{figure}

\section{Summary}
\label{sec4}
The EicC is a proposed electron-ion collider based on the High Intensity heavy-ion Accelerator Facility in China, aiming at the precision exploration of the partonic structure of nucleon/nucleus. In this work, we explore its sensitivities to the dark photons and ALPs coupling exclusive to photons. In both cases, the new particles are produced via the electron scattering with the photon cloud surrounding the ion, as shown in the left panel of Figs.~\ref{fig:Feyn_dp} and~\ref{fig:alp_diagram}. For light dark photons or ALPs, the production is dominated by coherent scattering, where the momentum transferred to the ion is smaller than the inverse of the ion's radius, leading to an enhancement in cross section proportional to $Z^2$. 

The dark photons are produced with a boost factor of approximately $E_e/m_{A^\prime} = 3.5 \times(1~\text{GeV}/m_{A^\prime})$, moving collinearly with the direction of the initial electron, as shown in Fig.~\ref{fig:8}. However, compared to the search at EIC~\cite{Balkin:2023gya}, the dark photons are produced more centrally and with lower energy. As a result, the lepton pair from their decay can be detected more efficiently, as illustrated in Fig.~\ref{fig:9}. The EicC has a larger center-of-mass energy and requires a shorter decay length of dark photons in the lab frame, compared to the traditional beam dump experiments. Consequently, the EicC can probe dark photons with stronger couplings and heavier masses, bridging the gap between beam dump experiments and high-energy colliders as shown in Fig.~\ref{fig:10}.   

The ALPs are produced via the photon fusion process. Due to the ion form factor, the photons from ALPs decay are more aligned with the electron as shown in Fig. \ref{fig:4}. Fig.~\ref{fig:5} demonstrates that the EicC has an advantage in probing ALPs lighter than 1 GeV, thanks to its high photon detection efficiency and low photon energy threshold. 
Furthermore, due to its lower center-of-mass energy compared to the EIC, the ALPs produced at the EicC are less boosted, which allows the EicC to detect long-lived ALPs more efficiently through displaced vertex searches, as shown in Fig.~\ref{fig:6}. The EicC demonstrates significant potential for detecting ALPs compared to existing experiments, both in prompt searches and displaced vertex searches.  

\begin{acknowledgments}
We thank Ye Tian and Weizhi Xiong for their helpful discussions on the detector performance. T.M is partly supported by the Yan-Gui Talent Introduction Program (grant No. 118900M128) and Chinese Academy of Sciences Pioneer Initiative "Talent Introduction Plan". H.L. is supported by Azrieli foundation and by the U.S. Department of
Energy under Grant Contract DE-SC0012704. 
\end{acknowledgments}
\appendix
\section{dark photon calculation}
\label{apppendB}
The differential cross section of coherent massive dark photon production depends on the the center-of-mass energy of incident particles $\sqrt{s}$ and four independent kinematic variables:
\begin{equation}
    m_{e A^{\prime}}^2 \equiv\left(p_e+p_{A^{\prime}}\right)^2,\quad t \equiv-\left(p_N-k_N\right)^2, \quad \mathrm{d} \Omega_{A^{\prime}} \equiv \mathrm{d} \cos \theta_{A^{\prime}} \mathrm{d} \varphi_{A^{\prime}},
\end{equation}
which are all defined in the center-of-mass frame of outgoing electron and dark photon. The differential cross section of dark photon coherent production is
\begin{equation}
    \frac{\mathrm{d} \sigma\left(e N \rightarrow e N A^{\prime}\right)}{\mathrm{d} m_{e A^{\prime}} \mathrm{d} t \mathrm{~d} \Omega_{A^{\prime}}}=\frac{e^6 \epsilon^2 p_{\mathrm{cm}}\left(m_{e A^{\prime}}, m_{A^{\prime}}, m_e\right)}{(2 \pi)^4 2^6 p_{\mathrm{cm}}^2\left(\sqrt{s}, m_e, m_N\right) s}\left(\frac{F(t)}{t}\right)^2 \mathcal{A} .
\end{equation}
where $p_{\mathrm{cm}}\left(x, y, z\right) \equiv \sqrt{\lambda\left(x^2, y^2, z^2\right)} /(2 x) $, $\lambda(x,y,z)=x^2+y^2+z^2-2xy-2xz-2yz$, and $F(t)$ is the elastic nuclear form factor\cite{Bjorken:2009mm} :
\begin{equation}
    F(t)=Z\left(\frac{1}{1+t / d_{pb^{208}}}\right),
\end{equation}
$\text { with } d_{pb^{208}}=A^{-2 / 3} 0.164 \mathrm{GeV}^2 \approx 0.00467 \mathrm{GeV}^2$. The amplitude \cite{Liu:2017htz} is
\begin{equation}
    \begin{aligned}
\mathcal{A} & =2\left(\frac{\tilde{s}^2+\tilde{u}^2}{\tilde{s} \tilde{u}}\right)\left(t+4 m_N^2\right)-\frac{8 t}{\tilde{s} \tilde{u}}\left(P \cdot k_e\right)^2-\frac{8 t}{\tilde{s} \tilde{u}}\left(P \cdot p_e\right)^2 \\
& -\frac{8 t}{\tilde{s} \tilde{u}} \frac{t_1+m_{A^{\prime}}^2}{2}\left(t+4 m_N^2\right)+2 \frac{(\tilde{s}+\tilde{u})^2}{\tilde{s}^2 \tilde{u}^2}\left(m_{A^{\prime}}^2+2 m_e^2\right)\left(t+4 m_N^2\right) t \\
& -8 \frac{(\tilde{s}+\tilde{u})^2}{\tilde{s}^2 \tilde{u}^2}\left(m_{A^{\prime}}^2+2 m_e^2\right)\left(\frac{\tilde{u} P \cdot k_e+\tilde{s} P \cdot p_e}{\tilde{s}+\tilde{u}}\right)^2,
\end{aligned}
\end{equation}
 with
 \begin{equation}
     \tilde{s} \equiv 2 p_e \cdot p_{A^{\prime}}+m_{A^{\prime}}^2, \quad \tilde{u} \equiv-2 k_e \cdot p_{A^{\prime}}+m_{A^{\prime}}^2, \quad P \equiv p_N+k_N, \quad t_1 \equiv-2 k_e \cdot p_e+2 m_e^2 .
 \end{equation}
The phase space integration limits are 
\begin{equation}
    -1<\cos \theta_{A^{\prime}}<1, \quad 0<\varphi_{A^{\prime}}<2 \pi, \quad m_e+m_{A^{\prime}}<m_{e A^{\prime}}<m_{e A^{\prime}}^{\max }(t), \quad t_{-}<t<t_{+},
\end{equation}
with
\begin{equation}
    m_{e A^{\prime}}^{\max }(t)=\sqrt{s+m_N^2-\frac{E_{\mathrm{cm}}\left(\sqrt{s}, m_N, m_e\right)\left(2 m_N^2+t\right)-p_{\mathrm{cm}}\left(\sqrt{s}, m_N, m_e\right) \sqrt{t\left(4 m_N^2+t\right)}}{m_N^2 / \sqrt{s}}},
\end{equation}
defining $E_{\mathrm{cm}}\left(x, y, z\right) \equiv\left(x^2+y^2-z^2\right) /(2 x)$. Lastly, the kinematic integration limit for $t$ is given by
\begin{equation}
\begin{aligned}
& \frac{t_{ \pm}}{2} \equiv E_{\mathrm{cm}}\left(\sqrt{s}, m_N, m_e\right) E_{\mathrm{cm}}\left(\sqrt{s}, m_N, m_e+m_{A^{\prime}}\right) \\
& \quad \pm p_{\mathrm{cm}}\left(\sqrt{s}, m_N, m_e\right) p_{\mathrm{cm}}\left(\sqrt{s}, m_N, m_e+m_{A^{\prime}}\right)-m_N^2
\end{aligned}
\end{equation}

\section{Resolution parameters of detector in the EicC}
\label{appendixA}
To compute $\Delta m_{\gamma\gamma}$, we simulate the detector effects on the kinematic variables of the final-state photons, including their energy $E$, polar angle $\theta$, and azimuthal angle $\phi$. The resolution effects on these variables $\sigma_E$, $\sigma_{\theta}$, and $\sigma_{\phi}$ are modeled by the following empirical formulas
\bea
\sigma_E(\eta,m_{\gamma\gamma}) &=& \frac{c_1^2}{m_{\gamma\gamma}} + c_2^2,\\
\sigma_{\theta}(\eta,m_{\gamma\gamma}) &=& d_1 \cdot e^{d_2 \cdot m_{\gamma\gamma} + d_3} + d_4,\\
\sigma_{\phi}(\eta, m_{\gamma\gamma}) &=& f_1 \cdot e^{f_2 \cdot m_{\gamma\gamma} + f_3} + f_4.
\eea
 These resolutions depend on the invariant mass of the two photons $m_{\gamma\gamma}$ and the pseudorapidity $\eta$ of the particles. The parameters $c_i$, $d_j$, and $f_j$ ($i=1,2$; $j=1,2,3,4$) are constants derived from experimental calibration data, which quantify the performance of the detector in measuring each kinematic variable. Their values are listed in Table~\ref{tb Resolution Parameters} for reference, which is provided by the EicC experimental group~\cite{EicC}. Using these resolutions, we simulate the detector smearing effects on the four-momenta of the photons. To determine the resolution $\Delta m_{\gamma\gamma}$, we fit the reconstructed $m_{\gamma\gamma}$ distribution with a Gaussian function and extract its standard deviation, which quantifies the spread of the distribution caused by detector effects. 

\begin{table}[H]
    \centering
    \begin{tabular}{ccccccccccc}
        \toprule
        $\eta$ & -2.7 & -2.1 & -1.5 & -0.9 & -0.3 & 0.3 & 0.9 & 1.5 & 2.1 & 2.7 \\ 
        \midrule
        $c_1$ & 1.566  & 0.755  & 0.694  & 4.122  & 4.563  & 4.636  & 4.630  & 4.729  & 4.378  & 5.713  \\ 
        $c_2$ & 1.001  & 1.442  & 1.381  & 2.514  & 1.406  & 1.213  & 1.328  & 1.786  & 1.794  & $-1.5\times10^{-6}$ \\ 
        $d_1$ & 0.074  & 0.071  & 0.068  & 0.078  & 0.097  & 0.097  & 0.079  & 0.061  & 0.057  & 0.074  \\ 
        $d_2$ & -2.005  & -1.853  & -1.728  & -2.741  & -2.775  & -2.678  & -2.962  & -2.101  & -1.832  & -7.421  \\ 
        $d_3$ & 1.498  & 1.479  & 1.441  & 1.472  & 1.697  & 1.701  & 1.592  & 1.495  & 1.392  & 1.646  \\ 
        $d_4$ & 0.109  & 0.107  & 0.098  & 0.227  & 0.262  & 0.260  & 0.215  & 0.103  & 0.072  & 0.106  \\ 
        $f_1$ & 0.078  & 0.069  & 0.067  & 0.084  & 0.103  & 0.102  & 0.087  & 0.059  & 0.058  & 0.053  \\ 
        $f_2$ & -1.902  & -1.569  & -1.790  & -2.859  & -2.246  & -2.208  & -2.358  & -3.037  & -1.949  & -2.440  \\ 
        $f_3$ & 1.553  & 1.449  & 1.425  & 1.551  & 1.757  & 1.753  & 1.690  & 1.460  & 1.411  & 1.414  \\ 
        $f_4$ & 0.104  & 0.105  & 0.104  & 0.208  & 0.202  & 0.202  & 0.160  & 0.139  & 0.074  & 0.091  \\ 
        \bottomrule
    \end{tabular}
    \caption{The resolution parameters of detector in the EicC.}
    \label{tb Resolution Parameters}
\end{table}

\providecommand{\href}[2]{#2}\begingroup\raggedright\endgroup

\end{document}